\documentclass{fundam}

\usepackage{makeidx} 
\usepackage{setspace}

\usepackage{epsfig,wrapfig} \usepackage{amssymb,amsmath,latexsym}
\usepackage{subfigure,mathabx,graphicx}

\input xy \xyoption{all}

\newcommand{\abb}[3]{#1 \colon #2 \rightarrow #3}

\long\def\omitthis#1{\relax}

\begin{document}
\setcounter{page}{1}
\issue{XXI~(2009)}

\title{Matrix Graph Grammars with Application Conditions} 

\author{Pedro Pablo P\'erez Velasco, Juan de Lara\\
School of Computer Science \\
Universidad Aut\'onoma de Madrid\\    	
Ciudad Universitaria de Cantoblanco, 28049 - Madrid, Spain\\
\{pedro.perez, jdelara\}{@}uam.es
} \maketitle

\runninghead{P.~P.~P\'erez, J.~de Lara}{Matrix Graph Grammars}

\begin{abstract}
  In the Matrix approach to graph transformation we represent {\em
    simple} digraphs and rules with Boolean matrices and vectors, and
  the rewriting is expressed using Boolean operators only. In previous
  works, we developed analysis techniques enabling the study of the applicability 
  of rule sequences, their independence,
  state reachability and the minimal graph able to fire a sequence.

  In the present paper we improve our framework in two ways. First, we
  make explicit (in the form of a Boolean matrix) some negative
  implicit information in rules. This matrix (called {\em nihilation
    matrix}) contains the elements that, if present, forbid the
  application of the rule (i.e. potential dangling edges, or newly
  added edges, which cannot be already present in the simple digraph).
  Second, we introduce a novel notion of application condition, which
  combines graph diagrams together with monadic second order logic. 
  This allows for more flexibility and expressivity than previous approaches,
  as well as more concise conditions in certain cases.
  We demonstrate that these application conditions can be embedded into
  rules (i.e. in the left hand side and the nihilation matrix), and
  show that the applicability of a rule with arbitrary application
  conditions is equivalent to the applicability of a sequence of plain
  rules without application conditions. Therefore, the analysis of the
  former is equivalent to the analysis of the latter, showing that in
  our framework no additional results are needed for the study of
  application conditions.  Moreover, all analysis techniques
  of~\cite{JuanPP_1,JuanPP_2} for the study of sequences can
  be applied to application conditions.

\end{abstract}


\section{Introduction}\label{sec:intro}

Graph transformation~\cite{graGraBook,handbook} is becoming
increasingly popular in order to describe system behaviour due to its
graphical, declarative and formal nature. For example, it has been
used to describe the operational semantics of Domain Specific Visual
Languages (DSVLs)~\cite{JVLC}, taking the advantage that it is
possible to use the concrete syntax of the DSVL in the rules, which
then become more intuitive to the designer.

The main formalization of graph transformation is the so called
algebraic approach~\cite{graGraBook}, which uses category theory in
order to express the rewriting step. Prominent examples of this approach
are the double~\cite{DPO:handbook,graGraBook} and
single~\cite{SPO:handbook} pushout (DPO and SPO), which have developed
interesting analysis techniques, for example to check sequential and
parallel independence between pairs of
rules~\cite{graGraBook,handbook}, or to calculate critical
pairs~\cite{Heckel,Lambers}.

Frequently, graph transformation rules are equipped with {\em
  application conditions} (ACs)~\cite{AC:Ehrig,graGraBook,HeckelW95},
stating extra (i.e.  in addition to the left hand side) positive and
negative conditions that the host graph should satisfy for the rule to
be applicable.  The algebraic approach has proposed a kind of ACs with
predefined diagrams (i.e. graphs and morphisms making the condition)
and quantifiers regarding the existence or not of matchings of the
different graphs of the constraint in the host
graph~\cite{AC:Ehrig,graGraBook}. Most analysis techniques for plain
rules (without ACs) have to be adapted then for rules with ACs (see
e.g.~\cite{Lambers} for critical pairs with negative ACs). Moreover,
different adaptations may be needed for different kinds of ACs. Thus,
a uniform approach to analyse rules with arbitrary ACs would be very
useful.

In previous works~\cite{JuanPP_1,JuanPP_2,JuanPP_4,MGGBook}, we
developed a framework (Matrix Graph Grammars, MGGs) for the
transformation of {\em simple} digraphs. Simple digraphs and their
transformation rules can be represented using Boolean matrices and
vectors. Thus, the rewriting can be expressed using Boolean operators
only. One important point is that, as a difference from other
approaches, we explicitly represent the rule dynamics (addition and
deletion of elements), instead of only the static parts (rule pre- and
post-conditions). This fact gives an interesting viewpoint enabling
useful analysis techniques, such as for example checking independence
of a sequence of arbitrary length and a permutation of it, or to
obtain the smallest graph able to fire a sequence. On the theoretical
side, our formalization of graph transformation introduces concepts
from many branches of mathematics, like Boolean algebra, group theory,
functional analysis, tensor algebra and logics~\cite{MGGBook}.  This
wealth of available mathematical results opens the door to new
analysis methods not developed so far, like sequential independence and
explicit parallelism not limited to pairs of sequences, applicability,
congruence and reachability. On the practical side, the
implementations of our analysis techniques, being based on Boolean
algebra manipulations, are expected to have a good performance.

In this paper we improve the framework, by extending grammar rules
with a matrix (the {\em nihilation} matrix) that contains the edges
that, if present in the host graph, forbid rule application. These are
potential dangling edges and newly added ones, which cannot be added
twice, since we work with simple digraphs. This matrix, which can be
interpreted as a graph, makes explicit some implicit negative
information in the rule's pre-condition. To the best of our knowledge,
this idea is not present in any approach to graph transformation.

In addition, we propose a novel approach for graph constraints and
ACs, where the diagram and the quantifiers are not fixed. For the
quantification, we use a full-fledged formula using monadic second
order logic (MSOL)~\cite{Courcelle}. We show that once the match is
considered, a rule with ACs can be transformed into plain rules, by
adding the positive information to the left hand side, and the
negative in the nihilation matrix. This way, the applicability of a
rule with arbitrary ACs is equivalent to the applicability of one of
the sequences of plain rules in a set: analysing the latter is
equivalent to analysing the former.  Thus, in MGGs, there is no need
to extend the analysis techniques to special cases of ACs. Although we
present the concepts in the MGGs framework, many of these ideas are
applicable to other approaches as well.


\noindent {\bf Paper organization}. Section~\ref{sec:MGGs} gives an
overview of MGGs. Section~\ref{sec:AC} introduces our graph
constraints and ACs. Section~\ref{sec:AC_rules} shows how ACs can be
embedded into rules. Section~\ref{sec:AC_seq} presents the equivalence
between ACs and sequences.
Section~\ref{sec:related} compares with related work and
Section~\ref{sec:conclusions} ends with the conclusions. This paper is
an extension of~\cite{GTVC}.

\section{Matrix Graph Grammars}
\label{sec:MGGs}

\begin{figure}[htbp]
  \centering \subfigure{
    \includegraphics[scale =
    0.6]{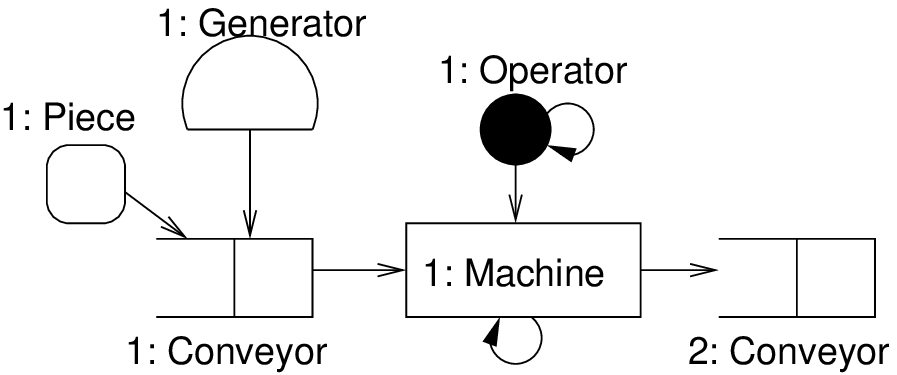}}\hspace{0.3cm} \subfigure{
    \includegraphics[scale = 0.7]{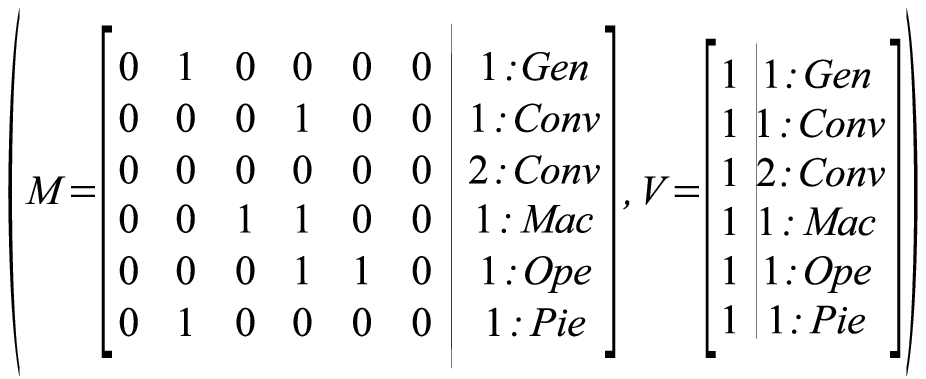}}
  \caption{Simple Digraph Example (left). Matrix Representation
    (right).}\label{fig:example_graph}
\end{figure}

{\bf Simple Digraphs.} We work with simple digraphs, which we
represent as $(M, V)$ where $M$ is a Boolean matrix for edges (the
graph {\em adjacency} matrix) and $V$ a Boolean vector for vertices or
nodes. We use the notation $|M|$ and $|V|$ to denote the set of edges
and nodes respectively.
Note that we explicitly represent the nodes of the graph with a
vector. This is necessary because in our approach we add and delete
nodes, and thus we mark the existing nodes with a $1$ in the
corresponding position of the vector. The left of Fig.~\ref{fig:example_graph}
shows a graph representing a production system made of a machine
(controlled by an operator), which consumes and produces pieces
through conveyors. Generators create pieces in conveyors. Self loops
in operators and machines indicate that they are busy.

Note that the matrix and the vector in the figure are the smallest
ones able to represent the graph. Adding zero elements to the vector
(and accordingly zero rows and columns to the matrix) would result in
equivalent graphs. Next definition formulates the representation of
simple digraphs.

\begin{definition}[Simple Digraph Representation]\label{def:simple_digraph}  
A simple digraph $G$ is represented by $G_M=(M, V)$ where $M$ is the
  graph's {\em adjacency} matrix and $V$ the Boolean vector of its
  nodes.
\end{definition}

\noindent {\bf Compatibility}. Well-formedness of graphs (i.e.,
absence of dangling edges) can be checked by verifying the identity
$\left\| \left( M \vee M^t \right) \odot \overline{V}\right\| _1 = 0$,
where $\odot$ is the Boolean matrix product (like the regular matrix
product, but with {\bf and} and {\bf or} instead of multiplication and
addition), $M^t$ is the transpose of the matrix $M$, $\overline{V}$ is
the negation of the nodes vector $V$, and $\| \cdot \|_1$ is an
operation (a norm, actually) that results in the {\bf or} of all the
components of the vector. We call this property {\em
  compatibility}~\cite{JuanPP_1}. Note that $M \odot \overline{V}$
results in a vector that contains a 1 in position $i$ when there is an
outgoing edge from node $i$ to a non-existing node. A similar
expression with the transpose of $M$ is used to check for incoming
edges.  The next definition formally characterizes compatibility.

\begin{definition}[Compatibility]\label{def:compatibility} A simple digraph $G_M=(M, V)$ is compatible iff $\left\| \left( M
      \vee M^t \right) \odot \overline{V}\right\| _1 = 0$.
\end{definition}

\noindent {\bf Typing}. A type is assigned to each node in $G=(M,V)$
by a function from the set of nodes $|V|$ to a set of types $T$,
$\abb{type}{|V|}{T}$. In Fig.~\ref{fig:example_graph} types are
represented as an extra column in the matrices, the numbers before the
colon distinguish elements of the same type. For edges we use the
types of their source and target nodes.

\begin{definition}[Typed Simple
  Digraph]\label{def:typed_simple_digraph}

  A typed simple digraph $G_T=(G_M, type)$ over a set of types $T$, is
  made of a simple digraph $G_M=(M, V)$, and a function from the set
  of nodes $|V|$ to the set of types $T$, $\abb{type}{|V|}{T}$.
\end{definition}

Next, we define the notion of partial morphism between typed simple
digraphs.

\begin{definition}[Typed Simple Digraph Morphism]\label{def:morphism}

  Given two simple digraphs $G_i=((M_i, V_i), \abb{type_i}{V_i}{T})$
  for $i=\{1, 2\}$, a morphism $\abb{f=(f_V, f_E)}{G_1}{G_2}$ is made
  of two partial injective functions $\abb{f_V}{|V_1|}{|V_2|}$,
  $\abb{f_E}{|M_1|}{|M_2|}$ between the set of nodes ($|V_i|$) and edges ($|M_i|$), s.t.
  $\forall v \in Dom(f_V), \: type_1(v)=type_2(f_V(v))$ and $\forall
  e=(n, m) \in Dom(f_E), \: f_E((n, m))=(f_V(n), f_V(m))$; where
  $Dom(f)$ is the domain of the partial function $f$.
\end{definition}

\noindent {\bf Productions.} A production, or rule, $p:L \rightarrow
R$ is a morphism of typed simple digraphs. Using a {\em static
  formulation}, a rule is represented by two typed simple digraphs
that encode the left and right hand sides (LHS and RHS). The matrices
and vectors of these graphs are arranged so that the elements
identified by morphism $p$ match (this is called completion, see
below).

\begin{definition}[Static Formulation of
  Production]\label{def:static_production}

  A production $p:L \rightarrow R$ is statically represented as $p=(
  L=(L^E, L^V, type^L); R=(R^E,$ $R^V, type^R))$, where $E$ stands for
  edges and $V$ for vertices.
\end{definition}

A production adds and deletes nodes and edges, therefore using a {\em
  dynamic formulation}, we can encode the rule's pre-condition (its
LHS) together with matrices and vectors representing the addition and
deletion of edges and nodes. We call such matrices and vectors $e$ for
``erase'' and $r$ for ``restock''.

\begin{definition}[Dynamic Formulation of
  Production]\label{def:dynamic_production}

  A production $p:L \rightarrow R$ is dynamically represented as $p=(
  L=(L^E, L^V, type^L); e^E, r^E; e^V,$ $r^V; type^r)$, where $type^r$
  contains the types of the new nodes, $e^E$ and $e^V$ are the
  deletion Boolean matrix and vector, $r^E$ and $r^V$ are the addition
  Boolean matrix and vector. They have a 1 in the position where the
  element is to be deleted or added respectively.
\end{definition}

The output of rule $p$ is calculated by the Boolean formula $R = p(L)
= r \vee \overline{e} \, L$, which applies both to nodes and edges
(the $\wedge$ ({\bf and}) symbol is usually omitted in formulae).

\begin{figure}[htbp]
  \centering \subfigure{
    \includegraphics[scale =
    0.5]{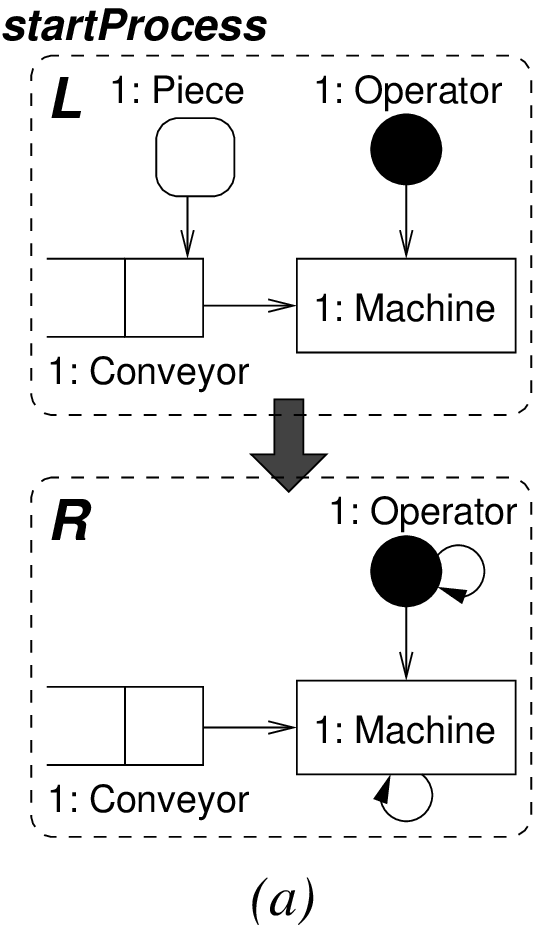}}\hspace{0.4cm} \subfigure{
    \includegraphics[scale = 0.65]{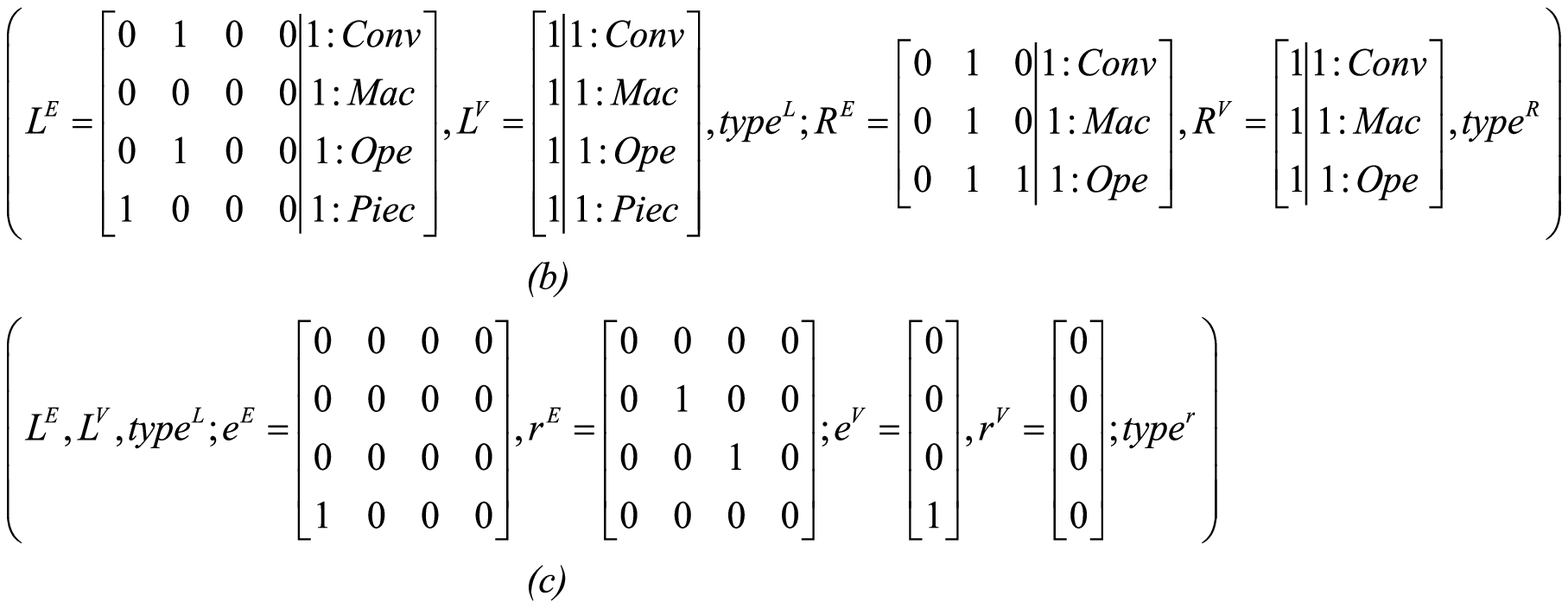}}
  \caption{(a) Rule Example. (b) Static Formulation. (c) Dynamic
    Formulation.}\label{fig:example_rule}
\end{figure}

\noindent {\bf Example.} Fig.~\ref{fig:example_rule} shows a rule and
its associated matrices. The rule models the consumption of a piece by
a machine.  Compatibility of the resulting graph must be ensured, thus
the rule cannot be applied if the machine is already busy, as it would
end up with two self loops, which is not allowed in a simple digraph.
This restriction of simple digraphs can be useful in this kind of
situations, and acts like a built-in negative AC. Later we will see
that the {\em Nihilation matrix} takes care of this restriction.

\noindent {\bf Completion.} In order to operate with the matrix representation of graphs of different
sizes, an operation called completion adds extra rows and columns with
zeros to matrices and vectors and rearranges rows and columns so
that the identified edges and nodes of the two graphs match.  For
example, in Fig.~\ref{fig:example_rule}, if we need to operate $L^E$ and $R^E$, 
completion adds a fourth 0-row and fourth 0-column to $R^E$. 

Stated in another way, whenever we have to operate graphs $G^1$ and
$G^2$, a morphism $\abb{f}{G^1}{G^2}$ (i.e. a partial function) has to
be defined. Completion rearranges the matrices and vectors of both
graphs so that the elements in $Dom(f)$ end up in the same row and column
of the matrices. Thus, after the
completion we have that $G^1 \wedge G^2 \cong Dom(f)$.  In the
examples, we omit such operation, assuming that matrices are completed
when necessary. Later we will operate with the matrices of different
productions, thus we have to select the elements (nodes and edges) of
each rule that get identified to the same element in the host graph.
That is, one has to establish morphisms between the LHS and RHS of the
different rules, and completion rearranges the matrices according to
the morphisms. Note that there may be different ways to complete two matrices,
by chosing different orderings for its rows and columns. 
This is because a simple digraph can be represented by many 
adjacency matrices, which differ in the order of rows and columns. In any case, the graphs
represented by the matrices are the same.

\noindent{\bf Nihilation Matrix.} In order to consider the elements in
the host graph that disable a rule application, we extend the notation
for rules with a new graph $N$. Its associated matrix $N^E$ specifies the
two kinds of forbidden edges: those incident to nodes which are going
to be erased and any edge added by the rule (which cannot be added
twice, since we are dealing with simple digraphs). Notice however
that $N^E$ considers only potential dangling edges with source and
target in the nodes belonging to $L^V$.

\begin{definition}[Nihilation Matrix]\label{def:nihilation_matrix}

  Given the production $p=( L=(L^E, L^V, type^L); e^E, r^E; e^V,$
  $r^V; type^r)$, its nihilation matrix $N^E$ contains non-zero
  elements in positions corresponding to newly added edges, and to
  non-deleted edges adjacent to deleted nodes.
\end{definition}

We extend the rule formulation with this nihilation matrix. The
concept of rule remains unaltered because we are just making explicit
some implicit information. Matrices are derived in the following
order: $\left( L, R \right) \mapsto \left( e, r \right) \mapsto N^E$.
Thus, a rule is \emph{statically} determined by its LHS and RHS $p =
\left( L, R \right)$, from which it is possible to give a dynamic
definition $p = \left(L; e, r \right)$, with $e=L\overline{R}$ and
$r=R\overline{L}$, to end up with a full specification including its
\emph{environmental} behaviour $p = \left(L, N^E; e, r \right)$.  No
extra effort is needed from the grammar designer, because $N^E$ can be
automatically calculated as the image by rule $p$ of a certain matrix
(see proposition~\ref{lemma:nihilMatrix}).

\begin{definition}[Full Dynamic Formulation of
  Production]\label{def:full_dynamic_production}

  A production $p:L \rightarrow R$ is dynamically represented as $p=(
  L=(L^E, L^V, type^L); N^E; e^E, r^E;$ $e^V, r^V; type^r)$, where
  $N^E$ is the nihilation matrix, $e^E$ and $e^V$ are the deletion
  Boolean matrix and vector, and $r^E$ and $r^V$ are the addition
  Boolean matrix and vector.
\end{definition}

Next proposition shows how to calculate the nihilation matrix using
the production $p$, by applying it to a certain matrix.

\begin{proposition}[Nihilation matrix]\label{lemma:nihilMatrix}

  The nihilation matrix $N^E$ of a given production $p$ is calculated
  as $N^E = p \left(\overline{D} \right)$ with $D = \overline{e^V}
  \otimes \overline{e^V}^t$. \footnote{Symbol $\otimes$ denotes the
    tensor product, which sums up the covariant and contravariant
    parts and multiplies every element of the first vector by the
    whole second vector.}
\end{proposition}

\noindent\emph{Proof.} Matrix $\overline{D}$ specifies potential dangling
edges incident to nodes in $p$'s LHS:
\begin{equation}
  \overline{D} = d^i_{\!j} = \left\{
    \begin{array}{ll}
      1 & \qquad if \; (e^V)^i = 1 \; or \; (e^V)^j = 1. \\
      0 & \qquad otherwise.
    \end{array} \right.
\end{equation}
Note that $D = \overline{e^V} \otimes \overline{e^V}^t$. Every
incident edge to a node that is deleted becomes dangling, except those
explicitly deleted by the production. In addition, edges added by the
rule cannot be present in the host graph, $N^E = r^E \vee
\overline{e^E} \left( \overline{D} \right) = p \left( \overline{D}
\right)$. $\blacksquare$


\noindent {\bf Example.} The nihilation matrix $N^E$ for the example
rule of Fig.~\ref{fig:example_rule} is calculated as follows:

\begin{equation}
  \overline{\overline{e^V} \otimes \overline{\left( e^V \right)}^{\,
      t}} = 
  \overline{ \left[
      \begin{array}{c}
        1 \\
        1 \\
        1 \\
        0 \\
      \end{array}
    \right] \otimes
    \left[
      \begin{array}{c}
        1 \\
        1 \\
        1 \\
        0 \\
      \end{array}
    \right]^t} =
  \left[
    \begin{array}{cccc}
      0 & 0 & 0 & 1 \\
      0 & 0 & 0 & 1 \\
      0 & 0 & 0 & 1 \\
      1 & 1 & 1 & 1 \\
    \end{array}
  \right]
  \nonumber
\end{equation}

The nihilation matrix is then given by :
\begin{equation}
  N^E = r \vee \overline e \overline{D} =
  \left[
    \begin{array}{cccc}
      0 & 0 & 0 & 0\\
      0 & 1 & 0 & 0\\
      0 & 0 & 1 & 0\\
      0 & 0 & 0 & 0\\
    \end{array}
  \right] \vee
  \overline {\left[
      \begin{array}{cccc}
        0 & 0 & 0 & 0\\
        0 & 0 & 0 & 0\\
        0 & 0 & 0 & 0\\
        1 & 0 & 0 & 0 \\
      \end{array} 
    \right]}
  \left[
    \begin{array}{cccc}
      0 & 0 & 0 & 1 \\
      0 & 0 & 0 & 1 \\
      0 & 0 & 0 & 1 \\
      1 & 1 & 1 & 1 \\
    \end{array} 
  \right] =
  \left[
    \begin{array}{cccc}
      0 & 0 & 0 & 1 \\
      0 & 1 & 0 & 1 \\
      0 & 0 & 1 & 1 \\
      0 & 1 & 1 & 1 \\
    \end{array}
  \right] \nonumber
\end{equation}

The matrix indicates any dangling edge from the deleted piece (the
edge to the conveyor is not signaled as it is explicitly deleted), as
well as self-loops in the machine and in the operator.

\begin{wrapfigure}{r}{0.2\textwidth}
  \vspace{-0.5cm}
  \centering
  \includegraphics[width=0.18\textwidth]{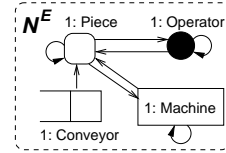}
  \caption{$N^E$ Graph for {\em startProcess}.}
  \label{fig:nihilation_example}
  \vspace{-0.5cm}
  \noindent 
\end{wrapfigure}
Matrix $N^E$ can be extended to a simple digraph by taking the nodes in the
LHS: $N=(N^E, L^V)$. Note that it defines a simple digraph, as
one basically needs to add the source and target nodes of the edges in $N^E$, which are a subset of the nodes
in $L^V$, because for the calculation of $N^E$ we have used the edges stemming from the nodes in $L^V$. 
Fig.~\ref{fig:nihilation_example} shows the
graph representation for the nihilation matrix of previous example.
The nihilation matrix should not be confused with the notion of {\em
  Negative Application Condition} (NAC)~\cite{graGraBook}, which is an
additional graph specified by the designer (i.e. not derived from the
rule) containing extra negative conditions. $\blacksquare$

The evolution of the rule's LHS (i.e. how it is transformed into the
RHS) is given by the production itself ($R = p(L) = r \vee
\overline{e} \, L$). It is interesting to analyse the behaviour of the
nihilation matrix, which is given by the next proposition.

\begin{proposition}[Evolution of the Nihilation
  Matrix]\label{th:Nevolution}

  Let $p:L \rightarrow R$ be a compatible production with nihilation
  matrix $N^E$. Then, the elements that must not appear once the
  production is applied are given by $p^{-1} \left( N^E \right)$, where
  $p^{-1}$ is the inverse of $p$ (the production that adds what $p$ deletes
  and vice versa, obtained by swapping $e$ and $r$).
\end{proposition}

\noindent \emph{Proof.} The elements that should not appear in the RHS
are potential dangling edges and those deleted by the production: $e
\vee \overline{D}$. This coincides with $p^{-1}(N^E)$ as shown by the
following set of identities:
\begin{equation}
  \label{eq:3}
  p^{-1} \left( N^E \right) = e \vee \overline{r} \,N^E = e \vee
  \overline{r} \left( r \vee \overline{e} \, \overline{D} \right) = e
  \vee \overline{e} \, \overline{r} \, \overline{D} = e \vee \overline{r}
  \, \overline{D} = e \vee \overline{D}.
\end{equation}
In the last equality of \eqref{eq:3} compatibility has been used,
$\overline{r} \, \overline{D} = \overline{D}$. $\blacksquare$

\noindent \textbf{Remark}. Though strange at a first glance, a dual
behaviour of the negative part of a production with respect to the
positive part should be expected. The fact that $N^E$ uses $p^{-1}$
rather than $p$ for its evolution is quite natural. When a production
$p$ erases one element, it asks its LHS to include it, so it demands
its presence. The opposite happens when $p$ adds some element. For
$N^E$ things happen in the opposite direction. If the production asks
for the addition of some element, then the size of $N^E$ (its number of
edges) is increased
while if some element is deleted, $N^E$ shrinks.

\begin{figure}[htbp]
  \centering
  \includegraphics[scale = 0.45]{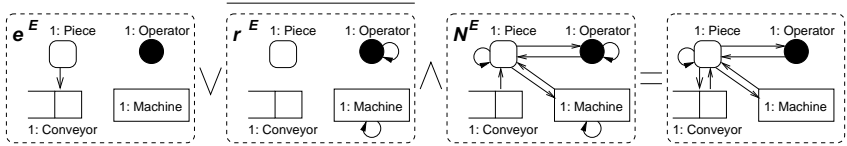}
  \caption{Evolution of Nihilation Matrix.}
  \label{fig:nihilation_evol}
\end{figure}

\noindent {\bf Example.} Fig.~\ref{fig:nihilation_evol} shows the
calculation of $startProcess^{-1}(N^E)$ using the graph representation
of the matrices in equation~\ref{eq:3}. $\blacksquare$

Next definition introduces a functional notation for rules (already
used in~\cite{JuanPP_2}), inspired by the Dirac or bra-ket
notation~\cite{braket}. This notation will be useful for reasoning and proving
the propositions in Section~\ref{sec:AC_seq}.

\begin{definition}[Functional Formulation of
  Production]\label{def:functional_rule}

  A production $p:L \rightarrow R$ can be depicted as
  $R=p(L)=\left\langle L, p\right\rangle$, splitting the static part
  (initial state, $L$) from the dynamics (element addition and
  deletion, $p$).
\end{definition}

Using such formulation, the {\em ket} operators (i.e. those to the
right side of the bra-ket) can be moved to the {\em bra} (i.e. left
hand side) by using their adjoints (which are usually decorated with
an asterisk).  We make use of this notation in
Section~\ref{sec:AC_seq}.

\noindent {\bf Match and Derivations.} Matching is the operation of
identifying the LHS of a rule inside a host graph (we consider only
injective matches). Given rule $p:L \rightarrow R$ and a simple
digraph $G$, any total injective morphism $m:L \rightarrow G$ is a
match for $p$ in $G$, thus it is one of the ways of {\em completing}
$L$ in $G$. The following definition considers not only the elements
that should be present in the host graph $G$ (those in $L$) but also
those that should not (those in the nihilation matrix, $N^E$).

\begin{definition}[Direct Derivation]\label{def:directDerivationDef}

  Given rule $p:L \rightarrow R$ and graph $G=(G^E, G^V)$ as in
  Fig.~\ref{fig:matches}(a), $d = \left( p, m \right)$ -- with $m =
  \left( m_L, m^E_N \right)$ -- is called a direct derivation with
  result $H=p^* \left( G \right)$ if the following conditions are
  satisfied:
  \begin{enumerate}
  \item There exist total injective morphisms $m_L : L \rightarrow G$ and $m^E_N : N^E
    \rightarrow \overline{G^E}$.
  \item $m_L(n) = m^E_N(n)$, $\forall n \in L^V$.
  \item The match $m_L$ induces a completion of $L$ in $G$. Matrices
    $e$ and $r$ are then completed in the same way to yield $e^*$ and
    $r^*$. The output graph is calculated as $H=p^*(G)=r^* \vee
    \overline {e^*} G$.
  \end{enumerate}

\begin{figure}[htbp]
  \centering
  \includegraphics[scale = 0.45]{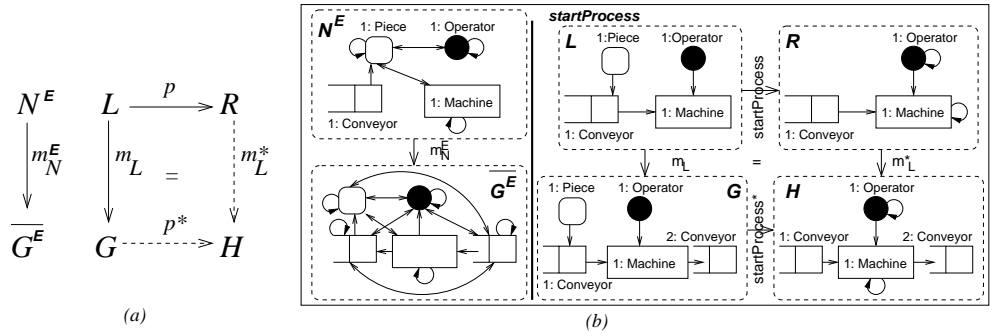}
  \caption{(a) Direct Derivation. (b) Example.} \label{fig:matches}
\end{figure}

\end{definition}

\noindent {\bf Remark.} 
Item 2 is needed to ensure that $L$ and $N^E$ are matched to
the same nodes in $G$.

\noindent {\bf Example.} Fig.~\ref{fig:matches}(b) shows the
application of rule {\em startProcess} to graph $G$. We have also
depicted the inclusion of $N^E$ in $\overline {G^E}$ (bidirectional
arrows have been used for simplification). $\overline {G^E}$ is the
complement (negation) of matrix $G^E$. $\blacksquare$

It is useful to consider the structure defined by the negation of the
host graph, $\overline G=(\overline {G^E}, \overline{G^V})$.  It is
made up of the graph $\overline {G^E}$ and the vector of nodes
$\overline{G^V}$. Note that the negation of a graph is not a graph
because in general compatibility fails, that is why the term
``structure'' is used.


The complement of a graph coincides with the negation of the adjacency
matrix, but while negation is just the logical operation, taking the
complement means that a completion operation has been performed
before. Hence, taking the complement of a matrix $G^E$ is the negation
with respect to some appropriate completion of $G$.  That is, the
complement of graph $G$ with respect to graph $A$, through a morphism
$\abb{f}{A}{G}$ is a two-step operation: (i) complete $G$ and $A$
according to $f$, yielding $G'$ and $A'$; (ii) negate $G'$.  As long
as no confusion arises negation and complements will not be
syntactically distinguished.

\begin{figure}[htbp]
  \centering
  \includegraphics[scale = 0.45]{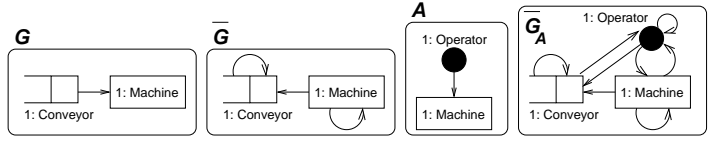}
  \caption{Finding Complement and negation of a Graph.}
  \label{fig:negation_graph}
\end{figure}

\noindent \textbf{Examples}. Suppose we have two graphs $A$ and $G$ as
those depicted in Fig.~\ref{fig:negation_graph} and that we want to
check that $A$ is not in $G$.  Note that $A$ is not contained in
$\overline G$ (an operator node does not even appear), but it does
appear in the negation of the completion of $G$ with respect to $A$
(graph $\overline{G}_{A}$ in the same figure).

In the context of Fig.~\ref{fig:matches}(b), we see that there is an
inclusion $startProcess^{-1}(N^E) \rightarrow \overline H$ (i.e. the
forbidden elements after applying production $startProcess$ are not in
$H$). This is so because we complete $H$ with an additional piece
(which was deleted from $G$). Note also that in Definition~\ref{def:directDerivationDef}, 
we have to complete $L$ and $G$ (step 3). As an occurrence of $L$ has to be found in $G$, 
all nodes of $L$ have to be
present in $G$ and thus $G$ is big enough to be able to find an inclusion $N^E \rightarrow \overline{G^E}$. 
$\blacksquare$

When applying a rule, {\em dangling edges} can occur. This is possible
because the nihilation matrix only considers dangling edges to nodes
appearing in the rule's LHS. However, a dangling edge can occur between
a node deleted by the rule and a node not considered by the rule's LHS.
In MGG, we
propose an SPO-like behaviour~\cite{JuanPP_1}, where the dangling
edges are deleted. Thus, if rule $p$ produces dangling edges (a fact
that is partially signaled by $m_N$) it is enlarged to explicitly
consider the dangling edges in the LHS. This is equivalent to adding a
pre-production (called $\varepsilon-$production) to be applied before
the original rule~\cite{JuanPP_2}.  Thus, rule $p$ is transformed into
sequence $p;p_{\varepsilon}$ (applied from right to left), where
$p_\varepsilon$ deletes the dangling edges and $p$ is applied as it
is. In order to ensure that both productions are applied to the same
elements (matches are non-deterministic), we defined a {\em marking}
operator $T_\mu$ which modifies the rules, so that the resulting rule
$T_\mu$($p_\varepsilon$), in addition, adds a special node connected
to the elements to be marked, and $T_\mu(p)$ in addition considers the
special node in the LHS and then deletes it. This is a technique to
control rule application by passing the match from one rule to the
next.

\noindent {\bf Analysis Techniques.}
In~\cite{JuanPP_1,JuanPP_2,JuanPP_4,MGGBook} we developed
some analysis techniques for MGGs, we briefly give an intuition to
those that will be used in Section~\ref{sec:analysisAC}.

One of the goals of our previous work was to analyse rule sequences
independently of a host graph. We represent a rule sequence as
$s_n=p_n; ...; p_1$, where application is from right to left (i.e.
$p_1$ is applied first). For its analysis, we {\em complete} the
sequence, by identifying the nodes across rules which are assumed to
be mapped to the same node in the host graph.

Once the sequence is completed, our notion of sequence {\em
  coherence}~\cite{JuanPP_1}~\cite{MGGCombinatorics}~\cite{MGGBook}
permits knowing if, for the given identification, the sequence is
potentially applicable (i.e. if no rule disturbs the application of
those following it). The formula for coherence results in a matrix and
a vector (which can be interpreted as a graph) with the problematic
elements. If the sequence is coherent, both should be zero, if not,
they contain the problematic elements. A coherent sequence is {\em
  compatible} if its application produces a simple digraph. That is,
no dangling edges are produced in intermediate steps.

Given a completed sequence, the minimal initial digraph (MID) is the
smallest graph that permits applying such sequence. Conversely, the
negative initial digraph (NID) contains all elements that should not
be present in the host graph for the sequence to be applicable. In
this way, the NID is a graph that should be found in $\overline G$ for
the sequence to be applicable (i.e. none of its edges can be found in
$G$).  If the sequence is not completed (i.e. no overlapping of rules
is decided), we can also give the set of all graphs able to fire such
sequence or spoil its application. We call them {\em initial digraph
  set} and {\em negative digraph set} respectively. See section~6
in~\cite{MGGCombinatorics} or sections~4.4 and~5.3 in~\cite{MGGBook}.

Other concepts we developed aim at checking sequential independence
(i.e. same result) between a sequence and a permutation of it. {\em
  G-Congruence} detects if two sequences (one permutation of the
other) have the same MID and NID. It returns two matrices and two
vectors, representing two graphs, which are the differences between
the MIDs and NIDs of each sequence respectively. Thus if zero, the
sequences have the same MID and NID. Two coherent and compatible
completed sequences that are G-congruent are sequential
independent. See section~7 in~\cite{MGGCombinatorics}
or section~6.1 in~\cite{MGGBook}.

\section{Graph Constraints and Application Conditions}
\label{sec:AC}

In this section, we present our concepts of {\em graph constraints}
(GCs) and {\em application conditions} (ACs).  A GC is defined as a
\emph{diagram} plus a MSOL formula.  The diagram is made of a set of
graphs and morphisms ({\em partial} injective functions) which specify
the relationship between elements of the graphs. The formula specifies
the conditions to be satisfied in order to make a host graph $G$
satisfy the GC (i.e. we check whether $G$ is a model for the diagram
and the formula). The domain of discourse of the formulae are simple
digraphs, and the diagram is a means to represent the interpretation
function \textbf{I}.\footnote{Recall that, in essence, the
  \emph{domain of discourse} is a set of individual elements which can
  be quantified over. The \emph{interpretation function} assigns
  meanings (semantics) to symbols~\cite{Logic}.}

GC formulae are made of expressions about graph inclusions. For this purpose,
we introduce the following two predicates:
\begin{eqnarray}
  P(X_1, X_2) = \forall m [F(m, X_1) \Rightarrow F(m, X_2)]\\
  Q(X_1, X_2) = \exists e [F(e, X_1) \wedge F(e, X_2)]
\end{eqnarray}

\noindent where predicate $F(m, X)$ states that element $m$ (a node or
an edge) is in graph $X$.  In this way, predicate $P(X_1, X_2)$ means
that graph $X_1$ is included in $X_2$.  Note that $m$ ranges over all
nodes and edges (edges are defined by their initial and final node) of
$X_1$, thus ensuring the containment of $X_1$ in $X_2$ (i.e.
preserving the graph structure).  Predicate $Q(X_1, X_2)$ asserts that
there is a partial morphism between $X_1$ and $X_2$, which is defined on at least
one edge. That is, $X_1$ and $X_2$ share an edge. In this case, $e$ ranges over all edges.

Predicates decorated with superindices $E$ or $V$ refer to {\em E}dges
or {\em V}ertices. Thus, $P^V(X_1, X_2)$ says that every {\em vertex}
in graph $X_1$ should also be present in $X_2$. Actually $P(X_1, X_2)$
is in fact a shortcut for stating that all vertices in $X_1$ should be
found in $X_2$ ($P^V(X_1, X_2)$), all edges in $X_1$ should be found
in $X_2$ ($P^E(X_1, X_2)$) and in addition the set of nodes found
should correspond to the source and target nodes of the
edges. 

Predicate $P(X_1, X_2)$ asks for an inclusion morphism $d_{12}:X_1 \hookrightarrow
X_2$. The diagram of the constraint may already include such morphism $d_{12}$ (i.e. the
diagram can be seen as a set of restrictions imposed on the
interpretation function \textbf{I}) and we can either permit extensions
of $d_{12}$ (i.e. the model -- host graph -- may relate more elements
of $X_1$ and $X_2$) or keep it as defined in the diagram. In this
latter case, the host graph should identify exactly the specified
elements in $d_{12}$ and keep different the elements not related by
$d_{12}$.
This is represented using predicate $P_U$, which can be expressed
using $P^E$:
\begin{eqnarray}\label{def:P_U}
  P^E_U(X_1,X_2) = \forall a [\neg(F(a,D) + F(a,coD))] = P^E(D,coD) \wedge P^E(D^C,coD^C)
\end{eqnarray}

\noindent where $D = Dom(d_{12})$, $coD = coDom(d_{12})$, ${}^C$
stands for the complement (i.e. $D^C$ is the complement of
$Dom(d_{12})$ w.r.t $X_1$) and $+$ is the \textbf{xor} operation.
A similar reasoning applies to nodes.

The notation (syntax) will be simplified by making the host graph $G$
the default second argument for predicates $P$ and $Q$.  Besides, it
will be assumed that by default total morphisms are demanded: unless
otherwise stated predicate $P$ is assumed.

\begin{wrapfigure}{r}{0.28\textwidth}
  \vspace{-0.4cm} \centering
  \includegraphics[width=0.25\textwidth]{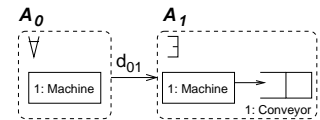}
  \caption{Diagram Example.} \label{fig:DiagExample}
\vspace{-0.4cm}
  \noindent
\end{wrapfigure}


\noindent {\bf Example}. Before starting with formal definitions, we
give an intuition of GCs.  The following GC is satisfied if for every
$A_0$ in $G$ it is possible to find a related $A_1$ in $G$: $\forall
A_0 \exists A_1 \left[ A_0 \Rightarrow A_1\right]$, equivalent by
definition to $\forall A_0 \exists A_1 \left[ P\!\!\left(A_0, G
  \right) \Rightarrow P\!\! \left(A_1, G\right) \right]$. Nodes and
edges in $A_0$ and $A_1$ are related through the diagram shown in
Fig.~\ref{fig:DiagExample}, which relates elements with the same
number and type. As a notational convenience, to enhance readability,
each graph in the diagram has been marked with the quantifier given in
the formula. If a total match is sought, no additional inscription is
presented, but if a partial match is demanded the graph is
additionally marked with a $Q$. Similarly, if a total match is
forbidden by the formula, the graph is marked with $\overline P$. This
convention will be used in most examples throughout the paper. The GC
in Fig.~\ref{fig:DiagExample} expresses that each machine should have
an output conveyor.$\blacksquare$

Note the identity $\overline{P}(A,G) = Q(A,\overline{G})$, which we
use throughout the paper. We take the convention that negations in
abbreviations apply to the predicate (e.g., $\exists A \left[
  \overline{A} \right] \equiv \exists A \left[ \overline{P} \left( A,
    G \right) \right]$) and not the negation of the graph's adjacency
matrix.

A bit more formally, the syntax of well-formed formulas is inductively
defined as in monadic second-order logic, which is first-order logic
plus variables for subsets of the domain of discourse. Across this
paper, formulas will normally have one variable term $G$ which
represents the host graph. Usually, the rest of the terms will be
given (they will be constant terms). Predicates will consist of $P$
and $Q$ and combinations of them through negation and binary
connectives. Next definition formally presents the notion of {\em
  diagram}.

\begin{definition}[Diagram]\label{def:diagram}
  A \emph{diagram} $\mathfrak{d}$ is a set of simple digraphs $\{A_i
  \}_{i \in I}$ and a set of partial injective morphisms $\{d_k \}_{k
    \in K}$  with $d_k: A_i \rightarrow A_j$.  Diagram $\mathfrak{d}$
  is well defined if every cycle of morphisms commute.
\end{definition}

The formulae in the constraints use variables in the set $\{A_i\}_{i
  \in I}$, and predicates $P$ and $Q$. Formulae are restricted to have
no free variables except for the default second argument of predicates
$P$ and $Q$, which is the host graph $G$ in which we evaluate the GC. Next definition
presents the notion of GC.

\begin{definition}[Graph Constraint]\label{def:graphConstraint}
  $GC = ( \mathfrak{d} = ( \{A_i \}_{i \in I}, \{ d_j \}_{j \in J} ),
  \mathfrak{f} ) $ is a graph constraint, where $\mathfrak{d}$ is a
  well defined diagram and $\mathfrak{f}$ a sentence with variables in
  $\{A_i\}_{i \in I}$. A constraint is called {\em basic} if $|I|=2$
  (with one bound variable and one free variable) and $J=\emptyset$.
\end{definition}

In general, there will be an outstanding variable among the $A_i$
representing the host graph, being the only free variable in
$\mathfrak{f}$. In previous paragraphs it has been denoted by $G$, the
default second argument for predicates $P$ and $Q$. We sometimes speak of
a ``GC defined over G''. A basic GC will be one made of just one graph
and no morphisms in the diagram (recall that the host graph is not
represented by default in the diagram nor included in the formulas).

Next, we define an AC as a GC where exactly one of the graphs in the
diagram is the rule's LHS (existentially quantified over the host
graph) and another one is the graph induced by the nihilation matrix
(existentially quantified over the negation of the host graph).

\begin{definition}[Application Condition]\label{def:weakPrecond}
  Given rule $p:L \rightarrow R$ with nihilation matrix $N^E$, an AC
  (over the free variable $G$) is a GC satisfying:
  \begin{enumerate}
  \item $\exists ! i,j$ such that $A_i = L$ and $A_j = N^E$.
  \item $\exists ! k$ such that $A_k = G$ is the only free variable.
  \item $\mathfrak{f}$ must demand the existence of $L$ in $G$ and the
    existence of $N^E$ in $\overline{G^E}$.
  \end{enumerate}
\end{definition}

The simple graph $G$ can be thought of as a host graph to which some
grammar rules are to be applied. For simplicity, we usually do not
explicitly show the condition 3 in the formulae of ACs, nor the
nihilation matrix $N^E$ in the diagram.  However, if omitted, both $L$
and $N^E$ are existentially quantified before any other graph of the
AC. Thus, an AC has the form $\exists L \nexists N^E ... [L \wedge
P(N^E, \overline G) \wedge ...]$.  Note the similarities between
Def.~\ref{def:weakPrecond} and that of derivation in
Def.~\ref{def:directDerivationDef}.

Actually, we can interpret the rule's LHS and its nihilation matrix as
the minimal AC a rule can have. Hence, any well defined production has
a natural associated AC.  Note also that, in addition to the AC
diagram, the structure of the rule itself imposes a relation between
$L$ and $N^E$ (and between $L$ and $R$).  For technical reasons,
related to converting pre- into post-conditions and viceversa, we
assume that morphisms in the diagram do not have codomain $L$ or
$N^E$. This is easily solved as we may always use their inverses due
to $d_i$'s injectiveness.

\noindent {\bf Semantics of Quantification.} In GCs or ACs, graphs are
quantified either existentially or universally. We now give the
intuition of the semantics of such quantification applied to basic
formulae.
Thus, we consider the four basic cases: (i) $\exists A[A]$, (ii)
$\forall A[A]$, (iii) $\nexists A[A]$, (iv) $\slash\!\!\forall A [A]$.

\begin{wrapfigure}{r}{0.35\textwidth}
  \centering
  \includegraphics[width=0.33\textwidth]{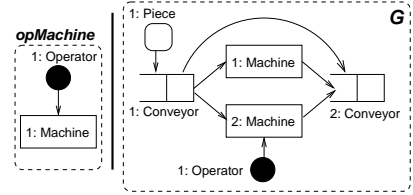}
  \caption{Quantification Example.} \label{fig:quantifier_semantics}
  \noindent
  \vspace{-1cm}
\end{wrapfigure}

Case (i) states that $G$ should include graph $A$.  For example,
in Fig.~\ref{fig:quantifier_semantics}, the GC $\exists
opMachine$ $[opMachine]$ demands an occurrence of $opMachine$ in $G$
(which exists).

Case (ii) demands that, for all {\em potential occurrences} of $A$ in
$G$, the shape of graph $A$ is actually found. The term potential
occurrences means all distinct maximal partial matches\footnote{A
  match is partial if it does not identify all nodes or edges of the
  source graph.  The domain of a partial match should be a graph.}
(which are total on nodes) of $A$ in $G$. A non-empty partial match in
$G$ is maximal, if it is not strictly included in another partial or
total match.  For example, consider the GC $\forall
opMachine[opMachine]$ in the
context of Fig.~\ref{fig:quantifier_semantics}.  There are two
possible instantiations of $opMachine$ (as there are two machines and
one operator), and these are the two input elements to the formula. As
only one of them satisfies $P(opMachine, G)$ -- the expanded form of
$[opMachine]$ -- the GC is not satisfied by $G$.


Case (iii) demands that, for all potential occurrences of $A$, none of
them should have the shape of $A$.  The term potential occurrence has
the same meaning as in case (ii). In
Fig.~\ref{fig:quantifier_semantics}, there are two potential
instantiations of the GC $\nexists opMachine[opMachine]$. As one of
them actually satisfies $P(opMachine, G)$, the formula is not
satisfied by $G$.

Finally, case (iv) is equivalent to $\exists A[\overline A]$, where by
definition $\overline A \equiv \overline P(A, G)$.  This GC states
that for all possible instantiations of $A$, one of them must not have
the shape of $A$. This means that a non-empty partial morphism $A
\rightarrow \overline G$ should be found.  The GC $\exists
opMachine[\overline{opMachine}]$ in
Fig.~\ref{fig:quantifier_semantics} is satisfied by $G$ because,
again, there are two possible instantiations, and one of them actually
does not have an edge between the operator and the machine.

Next definition formalizes the previous intuition, where we use the
following notation:
\begin{itemize}
\item $par^{max}(A, G) = \{ \abb{f}{A}{G} | f$ is a maximal non-empty
  partial morphism s.t. $Dom(f)^V=A^V \}$
\item $tot(A, G) = \{ \abb{f}{A}{G} | f$ is a total morphism $\}
  \subseteq par^{max}(A, G)$
\item $iso(A, G) = \{ \abb{f}{A}{G} | f$ is an isomorphism $\}
  \subseteq tot(A, G)$
\end{itemize}
\noindent where $Dom(f)^V$ are the nodes of the graph in the domain of
$f$.  Thus, $par^{max}(A, G)$ denotes the set of all potential
occurrences of a given constraint graph $A$ in $G$, where we require
all nodes in $A$ be present in the domain of $f$. Note that each $f
\in par^{max}$ may be empty in edges. 

\begin{definition}[Basic Constraint Satisfaction]\label{def:BasicGCSatisfied}

  \noindent The host graph $G$ satisfies $\exists A[A]$, written\footnote{The notation $G \models \mathfrak{f}$ is explained in more
  detail after Def.~\ref{def:GCSatisfied}.} $G \models \exists A[A]$ iff $\exists f \in par^{max}(A, G) \: [ f \in tot(A, G)]$.\\
  The host graph $G$ satisfies $\forall A[A]$, written\phantom{i} $G \models \forall A[A]$ iff $\forall f \in par^{max}(A, G) \: [f \in tot(A, G)]$.\\
\end{definition}

The diagrams associated to the formulas in previous definition have
been omitted for simplicity as they consist of a single element:
$A$. Recall that by default predicate $P$ is assumed as well as $G$ as
second argument, e.g. the first formula in previous definition
$\exists A[A]$ is actually $\exists A[P(A,G)]$. Note also that only these
two cases are needed, as one has $\nexists A[P(A,G)] \equiv \forall A[\overline P(A, G)]$
and $\slash\!\!\forall A[P(A,G)] \equiv \exists A[\overline P(A, G)] $.

Thus, this is a standard interpretation of MSOL formulae, save for the
domain of discourse (graphs) and therefore the elements of
quantification (maximal non-empty partial morphisms). Taking this fact
into account, next, we define when a graph satisfies an arbitrary $GC$.
This definition also applies to ACs.

\begin{definition}[Graph Constraint
  Satisfaction]\label{def:GCSatisfied}
  We say that $\mathfrak{d}_0 = (\{A_i\}, \{d_j\})$ satisfies the
  graph constraint $GC = (\mathfrak{d} = (\{X_i\},$$\{d_j\}),
  \mathfrak{f})$ under the interpretation function $I$, written $(I,
  \mathfrak{d}_0) \models \mathfrak{f}$, if $\mathfrak{d}_0$ is a model for
  $\mathfrak f$ that satisfies the element relations\footnote{As any
    mapping, $d_j$ assigns elements in the domain to elements in the
    codomain.  Elements so related should be mapped to the same
    element. For example, Let $a \in X_1$ and $d_{1i}:X_1 \rightarrow
    X_i$ with $b = d_{12}(a)$ and $c = d_{13}(a)$. Further, assume
    $d_{23}:X_2 \rightarrow X_3$, then $d_{23}(b) = c$.} specified by the
  diagram $\mathfrak d$, and the following interpretation for the
  predicates in $\mathfrak{f}$:
  \begin{enumerate}
  \item $I\left(P \left( X_i, X_j \right)\right) = m^T: X_i
    \rightarrow X_j$ total injective morphism.
  \item $I\left(Q \left( X_i, X_j \right)\right) = m^P: X_i
    \rightarrow X_j$ partial injective morphism, non-empty in edges.
  \end{enumerate}
  where $m^T \vert_D = d_k = m^P \vert_D$ with\footnote{It can be the
    case that $Dom\left(m^P \right) \cap Dom \left(d_k \right) =
    \emptyset$.} $d_k:X_i \rightarrow X_j$ and $D=Dom\left(d_k
  \right)$. The interpretation of quantification is as in
  Def.~\ref{def:BasicGCSatisfied} but setting $X_i$ and $X_j$ instead
  of $A$ and $G$, respectively.
\end{definition}

The notation deserves the following comments:
\begin{enumerate}
\item The notation $(I, \mathfrak{d}_0) \models \mathfrak{f}$ means
  that the formula $\mathfrak{f}$ is satisfied under interpretation
  given by $I$, assignments given by morphisms specified in
  $\mathfrak{d}_0$ and substituting the variables in $\mathfrak{f}$
  with the graphs in $\mathfrak{d}_0$.
\item As commented after Def.~\ref{def:graphConstraint}, in many cases
  the formula $\mathfrak{f}$ will have a single variable (the one
  representing the host graph $G$) and always the interpretation
  function will be that given in Def.~\ref{def:GCSatisfied}. We may
  thus write $G \models \mathfrak{f}$ which is the notation that
  appears in Def.~\ref{def:BasicGCSatisfied}. The notation $G
  \models GC$ may also be used.
\item Similarly, as an AC is just a GC where $L$, $N^E$ and $G$ are
  present, we may write $G \models AC$. For practical
  purposes, we are interested in testing whether, given a host graph
  $G$, a certain match $\abb{m_L}{L}{G}$ satisfies the AC. In this
  case we write $(G, m_L) \models AC$. In this way, the satisfaction
  of an AC by a match and a host graph is like the satisfaction of a
  GC by a graph $G$, where a morphism $m_L$ is already specified in the diagram of the
  GC.
\end{enumerate}


\noindent \textbf{Remark}. For technical reasons, we require all
graphs in the GC for which a partial morphism is demanded to be found
in the host graph to have at least one edge and be connected. That is
why $m^P$ has to be non-empty in edges.



\begin{figure}[htbp]
 \centering
 \includegraphics[scale = 0.5]{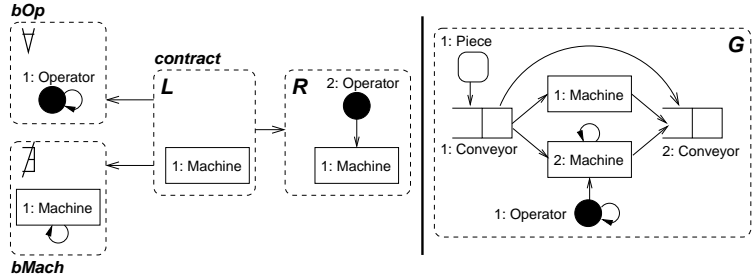}
 \caption{Satisfaction of Application Condition.}
 \label{fig:example_fulfill1}
\end{figure}
\noindent {\bf Examples.} Fig.~\ref{fig:example_fulfill1} shows rule
{\em contract}, with an AC given by the diagram in the figure (where
morphisms identify elements with the same type and number, this
convention is followed throughout the paper), together with formula
$\exists L \: \nexists bMach \: \forall bOp [L \wedge bMach \wedge
bOp]$. The rule creates a new operator, and assigns it to a machine.
The rule can be applied if there is a match of the LHS (a machine is
found), the machine is not busy ($\nexists bMach[bMach]$), and all
operators are busy ($\forall bOp[bOp]$). Graph $G$ to the right
satisfies the AC, with the match that identifies the machine in the
LHS with the machine in $G$ with the same number.

Using the terminology of ACs in the algebraic
approach~\cite{graGraBook}, $\nexists bMach[bMach]$ is a negative
application condition (NAC). On the other hand, there is no equivalent
to $\forall bOp[bOp]$ in the algebraic approach, but in this case it
could be emulated by a diagram made of two graphs stating that if an
operator exists then it does not have a self-loop. However, this is
not possible in all cases as next example shows.

\begin{figure}[htbp]
  \centering
  \includegraphics[scale = 0.42]{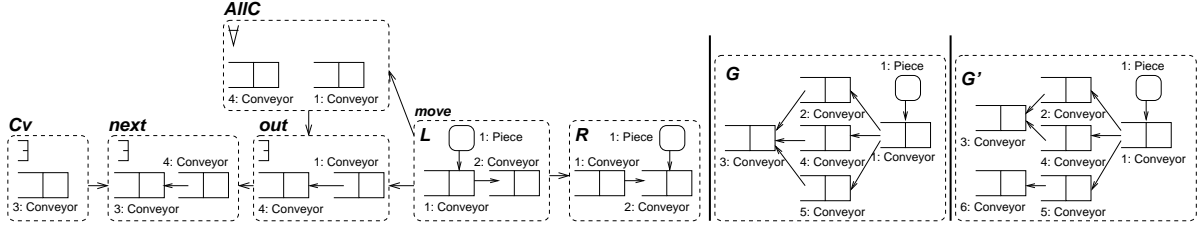}
  \caption{Example of Application Condition.} \label{fig:example_AC}
\end{figure}

Fig.~\ref{fig:example_AC} shows rule {\em move}, which has an AC with
formula: $\exists Cv \: \forall AllC \: \exists out \: \exists next [
(AllC \wedge out) \Rightarrow $ $( next \wedge Cv )]$. As previously
stated, in this example and the followings, the rule's LHS and the
nihilation matrix are omitted in the AC's formula.  The example AC
checks whether all conveyors connected to conveyor 1 in the LHS reach
a common target conveyor in one step. We can use ``global''
information, as graph $Cv$ has to be found in $G$ and then all output
conveyors are checked to be connected to it ($Cv$ is existentially
quantified in the formula before the universal). Note that we first
obtain all possible conveyors ($\forall AllC$).  As the
identifications of the morphism $L\rightarrow AllC$ have to be
preserved, we consider only those potential instances of $AllC$ with
$1: Conveyor$ equal to $1: Conveyor$ in $L$. From these, we take those
that are connected ($\exists out$), and which therefore have to be
connected with the conveyor identified by the LHS.  Graph $G$
satisfies the AC, while graph $G'$ does not, as the target conveyor
connected to $5$ is not the same as the one connected to $2$ and $4$.
To the best of our efforts it is not possible to express this
condition using the standard ACs in the DPO approach given
in~\cite{graGraBook}. $\blacksquare$

\section{Embedding Application Conditions into Rules}
\label{sec:AC_rules}

In this section, the goal is to embed arbitrary ACs into rules by
including the positive and negative coditions in $L$ and $N^E$
respectively.  It is necessary to check that direct derivations can be
the codomain of the interpretation function, that is, intuitively we
want to assert whether ``MGG + AC = MGG'' and ``MGG + GC = MGG''.

As stated in previous section, in direct derivations, the matching
corresponds to formula $\exists L \exists N^E$ $\left[ L \wedge
  P\left(N^E, \overline{G^E} \right) \right] $, but additional ACs may
represent much more general properties, due to universal quantifiers
and partial morphisms. Normally, plain rules (without ACs) in the
different approaches to graph transformation do not care about
elements that cannot be present. If so, a match is just $\exists L
[L]$.  Thus, we seek for a means to translate universal quantifiers
and partial morphisms into existential quantifiers and total
morphisms.

For this purpose, we introduce two operations on basic
diagrams:~\emph{closure} ($\mathfrak{C}$), dealing with universal
quantifiers only, and \emph{decomposition} ($\mathfrak{D}$), for
partial morphisms only (i.e. with the $Q$
predicate). 

The closure operator converts a universal quantification into a number
of existentials, as many as maximal partial matches there are in the
host graph (see definition~\ref{def:BasicGCSatisfied}).  Thus, given a
host graph $G$, demanding the universal appearance of graph $A$ in $G$
is equivalent to asking for the existence of as many replicas of $A$
as partial matches of $A$ are in $G$.

\begin{definition}[Closure]\label{def:closureDef}

  Given $GC = \left( \mathfrak{d}, \mathfrak{f} \right)$ with diagram
  $\mathfrak{d} = \{ A \}$, ground formula $\mathfrak{f} = \forall A
  [A]$ and a host graph $G$, the result of applying $\mathfrak{C}$ to
  $GC$ is calculated as follows:
  \begin{eqnarray}\label{eq:closure}
    \mathfrak{d} & \longmapsto & \mathfrak{d}' = \left(\{ A^1, \ldots,
      A^n \}, d_{ij}:A^i \rightarrow A^j \right) \nonumber \\
    \mathfrak{f} & \longmapsto & \mathfrak{f}' = \exists A^1 \ldots
    \exists A^n \left[ \bigwedge_{i=1}^n A^i \bigwedge_{i, j=1, j >
        i}^n P_U(A^i, A^j) \right]
  \end{eqnarray} 
  with $A^i \cong A$, $d_{ij} \not \in iso(A^i, A^j)$, $\mathfrak{C}
  \left( GC \right) = GC' = \left( \mathfrak{d}', \mathfrak{f}'
  \right)$ and $n=|par^{max}(A, G)|$.
    
\end{definition}

\noindent {\bf Remark.} Completion creates a morphism $d_{ij}$ between
each different $A^i$ and $A^j$ (both isomorphic to $A$), but morphisms are not needed in both
directions (i.e. $d_{ji}$ is not needed).  The condition that morphism
$d_{ij}$ must not be an isomorphism means that at least one element of
$A^i$ and $A^j$ has to be identified in different places of $G$. This
is accomplished by means of predicate $P_U$ (see its definition in
equation~\ref{def:P_U}), which ensures that the elements not related
by $\abb{d_{ij}}{A^i}{A^j}$, are not related in $G$.

The interpretation of the closure operator is that demanding the
universal appearance of a graph is equivalent to the existence of all
of its potential instances (i.e. those elements in $par^{max}$) in the
specified digraph ($G$, $\overline{G}$ or some other).  Some nodes can
be the same for different identifications ($d_{ij}$), so the procedure
does not take into account morphisms that identify every single node,
$d_{ij} \not \in iso( A^i, A^j )$.  Therefore, each $A^i$ contains the
image of a potential match of $A$ in $G$ (there are $n$ possible
occurrences of $A$ in $G$) and $d_{ij}$ identifies elements considered
equal.

\begin{figure}
  \centering
  \includegraphics[scale = 0.45]{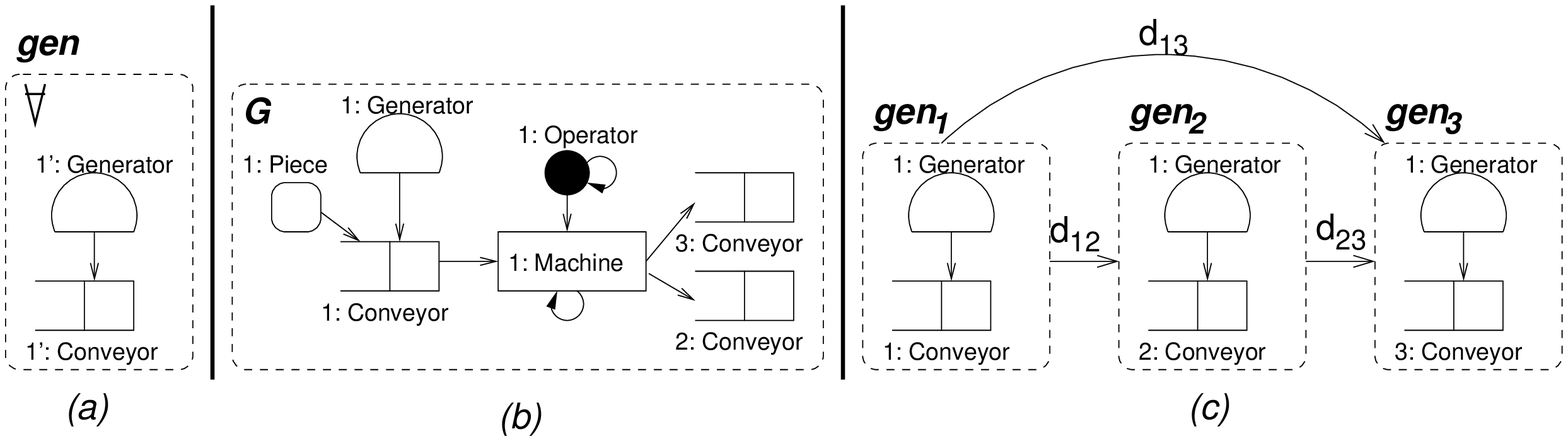}
  \caption{(a) GC diagram. (b) Graph where GC is to be evaluated. (c)
    Closure of GC w.r.t. G.} \label{fig:example_closure1}
\end{figure}

\noindent {\bf Example.} Assume the diagram to the left of
Fig.~\ref{fig:example_closure1}, made of just graph $gen$, together
with formula $\forall gen[gen]$, and graph $G$, where such GC is to be
evaluated. The GC asks $G$ for the existence of all potential
connections between each generator and each conveyor.  Performing
closure we obtain $\mathfrak{C}((gen, \forall
gen[gen]))=(\mathfrak{d}_C, \exists gen_1 \exists gen_2 \exists gen_3
[gen_1 \wedge gen_2 \wedge gen_3 \wedge P_U(gen_1, gen_2) \wedge
P_U(gen_1, gen_3) \wedge P_U(gen_2, gen_3) ])$, where diagram
$\mathfrak{d}_C$ is shown to the right of
Fig.~\ref{fig:example_closure1}, and each $d_{ij}$ identifies elements
with the same number and type. The closure operator makes explicit
that three potential occurrences must be found (as $|par^{max}(gen,
G)|=3$), thus, taking information from the graph where the GC is
evaluated and placing it in the GC itself.  $\blacksquare$

The idea behind decomposition is to split a graph into its basic
components to transform partial morphisms into total morphisms of one
of its parts.  For this purpose, the decomposition operator
$\mathfrak{D}$ splits a digraph $A$ into its edges, generating as many
digraphs as edges in $A$. As stated in remark 1 of
definition~\ref{def:GCSatisfied}, all graphs for which the GC asks for
a partial morphism are forbidden to have isolated nodes.  We are
more interested in the behaviour of edges (which to some extent
comprises nodes as source and target elements of the edges, except for
isolated nodes) than on nodes alone as they define the \emph{topology}
of the graph. This is also the reason why predicate $Q$ was defined to
be true in the presence of a partial morphism non-empty in edges.  If
so desired, in order to consider isolated nodes, it is possible to
define two decomposition operators, one for nodes and one for edges,
but this is left for future work.

\begin{definition}[Decomposition]\label{def:decompDef}

  Given $GC = \left( \mathfrak{d}, \mathfrak{f} \right)$ with ground
  formula $\mathfrak{f} = \exists A [Q(A)]$ and diagram $\mathfrak{d}
  = \{ A \}$, $\mathfrak{D}$ acts on $GC$ -- $\mathfrak{D} \left( GC
  \right) = AC' = \left( \mathfrak{d}', \mathfrak{f}' \right)$ -- in
  the following way:
  \begin{eqnarray}
    \mathfrak{d} & \longmapsto & \mathfrak{d}' = \left(\{ A^1, \ldots,
      A^n \}, d_{ij}:A^i \rightarrow A^j \right) \nonumber \\
    \mathfrak{f} & \longmapsto & \mathfrak{f}' = \exists A^1 \ldots
    \exists A^n \left[ \bigvee_{i=1}^n A^i \right] \label{eq:decomp}
  \end{eqnarray}
  with $n = \#\{edg(A)\}$, the number of edges of $A$, and $Q(A^i,
  A)$, where $A^i$ contains a single edge of $A$.
    
\end{definition}

Demanding a partial morphism is equivalent to asking for the existence
of a total morphism of some of its edges, that is, each $A^i$ contains
exactly one of the edges of $A$.

\begin{figure}[htbp]
  \centering
  \includegraphics[scale = 0.4]{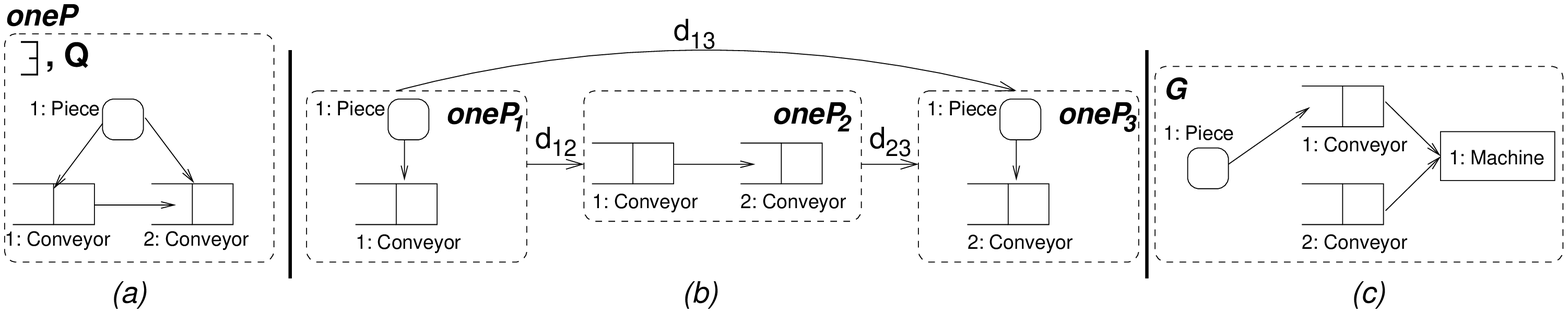}
  \caption{(a) GC diagram. (b) Decomposition of the GC. (c) Graph
    satisfying the GC. }\label{fig:example_decomp1}
\end{figure}

\noindent {\bf Example.} Consider $GC=(oneP, \exists oneP[Q(oneP)])$,
where graph $oneP$ is shown to the left of
Fig.~\ref{fig:example_decomp1}.  The constraint is satisfied by a host
graph $G$ if there is a partial morphism non-empty in edges
$\abb{m^P}{oneP}{G}$.  Thus, we require that either the two conveyors
are connected, or there is a piece in one of them.  Using
decomposition, we obtain $\mathfrak{D}(GC)=(\mathfrak{d}_D, \exists
oneP_1 \exists oneP_2 \exists oneP_3$ $[oneP_1 \vee oneP_2 \vee
oneP_3])$. Diagram $\mathfrak{d}_D$ is shown in
Fig.~\ref{fig:example_decomp1}(b), together with a graph $G$
satisfying the constraint in Fig.~\ref{fig:example_decomp1}(c).
Note that this constraint can be expressed more concisely than in
other approaches, like the algebraic/categorical one of~\cite{graGraBook}.

Note how, decomposition is not affected by the host graph to which it
is to be evaluated. Also, we do not care whether some graphs in the
decomposition are matched in the same place in the host graph (e.g.
$oneP_1$ and $oneP_3$), as the GC just requires one of them to be
found.  $\blacksquare$

Now we show the main result of this section, which states that it is
possible to reduce any formula in an AC (or GC) into another one using
existential quantifiers and total morphisms only. This theorem is of
interest because derivations as defined in MGGs (the matching part)
use only total morphisms and existential quantifiers.

\begin{theorem}[$\exists - P$ reduction]\label{th:embeddingGC}

  Let $GC = \left( \mathfrak{d}, \mathfrak{f}\left( P, Q \right)
  \right)$ with $\mathfrak{f}$ a ground formula, $\mathfrak{f}$ can be
  transformed into a logically equivalent $\mathfrak{f'} =
  \mathfrak{f}'(P)$ with existential quantifiers only.
\end{theorem}

\noindent \emph{Proof.} Let the depth of a graph for a fixed node
$n_0$ be the maximum over the shortest path (to avoid cycles) starting
in any node different from $n_0$ and ending in $n_0$. The depth of a graph
is the maximum depth for all its nodes.
Diagram
$\mathfrak{d}$ is a graph where nodes are digraphs $A_i$ and edges are
morphisms $d_{ij}$. We use $depth\left( GC \right)$ to denote the
depth of $\mathfrak{d}$. In order to prove the theorem we apply
induction on the depth, checking out every case. There are 16
possibilities for $depth\left(\mathfrak{d}\right) = 1$ and a single
element $A$, summarized in Table \ref{tab:possibilitiesSingleCase}.

\begin{table*}[hbtp]
  \centering
  \begin{tabular}{|rl|rl||rl|rl|}
    \hline
    \begin{Large}\phantom{I}\end{Large}(1) & $\exists A [ A ]$ & (5) & $\slash\!\!\forall A [\overline{A}]$ & (9) & $\exists A [\overline{Q}(A)]$ & (13) & $\slash\!\!\forall A [Q(A)]$\\
    \hline
    \begin{Large}\phantom{I}\end{Large}(2) & $\exists A [\overline{A}]$ & (6) & $\slash\!\!\forall A [A]$ & (10) & $\exists A [Q(A)]$ & (14) & $\slash\!\!\forall A [\overline{Q}(A)]$\\
    \hline
    \begin{Large}\phantom{I}\end{Large}(3) & $\nexists A [\overline{A}]$ & (7) & $\forall A [A]$ & (11) & $\nexists A [Q(A)]$ & (15) & $\forall A [\overline{Q}(A)]$\\
    \hline
    \begin{Large}\phantom{I}\end{Large}(4) & $\nexists A [ A ]$ & (8) & $\forall A [\overline{A}]$ & (12) & $\nexists A [\overline{Q}(A)]$ & (16) & $\forall A [Q(A)]$\\
    \hline
  \end{tabular}
  \caption{All Possible Diagrams for a Single Element.}
  \label{tab:possibilitiesSingleCase}
\end{table*}

Elements in the same row for each pair of columns are related using
equalities $\nexists A[A] = \forall A[\overline{A}]$ and
$\slash\!\!\forall A[A] = \exists A[\overline{A}]$, so it is possible
to reduce the study to cases (1)--(4) and (9)--(12). Identities
$\overline{Q}(A) = P(A, \overline{G})$ and $Q(A) = \overline{P}(A,
\overline{G})$ reduce (9)--(12) to formulae (1)--(4):

\begin{eqnarray}
  \exists A [\overline{Q}(A)] = \exists A\left[P(A, \overline{G})\right]
  &, & \exists A [Q(A)] = \exists A\left[\overline{P}(A, \overline{G})\right] \nonumber \\
  \nexists A [Q(A)] = \nexists A\left[\overline{P}(A, \overline{G})\right] &, &
  \nexists A [\overline{Q}(A)] = \nexists A\left[P(A, \overline{G})\right]. \nonumber
\end{eqnarray}

Thus, it is enough to study the first four cases, but we have to
specify if $A$ must be found in $G$ or $\overline{G}$. Finally, all
cases in the first column can be reduced to (1):
\begin{itemize}
\item (1) is the definition of match.
\item (2) can be transformed into total morphisms (case 1) using
  operator $\mathfrak{D}$: $\exists A\left[\overline{A}\right]
  =\exists A \left[Q(A,\overline{G})\right] = \exists A^1\ldots\exists
  A^n \left[\bigvee_{i=1}^n P\left(A^i, \overline{G}\right)\right]$.
\item (3) can be transformed into total morphisms (case 1) using
  operator $\mathfrak{C}$: $\nexists A\left[\overline{A}\right] =
  \forall A [A] = \exists A^1\ldots\exists A^n \left[\bigwedge_{i=1}^n
    A^i \right]$. Here for simplicity, the conditions on $P_U$ are
  assumed to be satisfied and thus have not been included.
\item (4) combines (2) and (3), where operators $\mathfrak{C}$ and
  $\mathfrak{D}$ are applied in order $\mathfrak{D}\circ\mathfrak{C}$
  (see remark below): $ \nexists A[A] = \forall A
  \left[\overline{A}\right] = \exists A^{11}\ldots\exists A^{mn}
  \left[\bigwedge_{i=1}^m \bigvee_{j=1}^n P\left(A^{ij},
      \overline{G}\right)\right]$.
\end{itemize}

If there is more than one element at depth 1, this same procedure can
be applied mechanically (well-definedness guarantees independence with
respect to the order in which elements are selected). Note that if
depth is 1, graphs on the diagram are unrelated (otherwise, depth $>$
1).

{\bf Induction Step.} When there is a universal quantifier $\forall
A$, according to equation \ref{eq:closure}, elements of $A$ are
replicated as many times as potential instances of $A$ can be found in
the host graph.  In order to continue the application procedure, we
have to clone the rest of the diagram for each replica of $A$, except
those graphs which are existentially quantified before $A$ in the
formula. That is, if we have a formula $\exists B \forall A \exists
C$, when performing the closure of $A$, we have to replicate $C$ as
many times as $A$, but not $B$.  Moreover $B$ has to be connected to
each replica of $A$, preserving the identifications of the morphism $B
\rightarrow A$.  More in detail, when closure is applied to $A$, we
iterate on all graphs $B_j$ in the diagram:

\begin{itemize}

\item If $B_j$ is existentially quantified after $A$ ($\forall A ...
  \exists B_j$) then it is replicated as many times as $A$.
  Appropriate morphisms are created between each $A^i$ and $B^i_j$ if
  a morphism $\abb{d}{A}{B}$ existed. The new morphisms identify
  elements in $A^i$ and $B^i_j$ according to $d$.  This permits finding
  different matches of $B_j$ for each $A^i$, some of which can be
  equal.\footnote{If for example there are three instances of $A$ in
    the host graph but only one of $B_j$, then the three replicas of
    $B$ are matched to the same part of $G$.}

\item If $B_j$ is existentially quantified before $A$ ($\exists B_j
  ... \forall A$) then it is not replicated, but just connected to
  each replica of $A$ if necessary. This ensures that a unique $B_j$
  has to be found for each $A^i$. Moreover, the replication of $A$ has
  to preserve the shape of the original diagram. That is, if there is
  a morphism $\abb{d}{B}{A}$, then each $\abb{d_i}{B}{A^i}$ has to
  preserve the identifications of $d$ (this means that we take only
  those $A^i$ which preserve the structure of the diagram).

\item If $B_j$ is universally quantified (no matter if it is
  quantified before or after $A$), again it is replicated as many
  times as $A$. Afterwards, $B_j$ will itself need to be replicated
  due to its universality. The order in which these replications are
  performed is not relevant as $\forall A \forall B_j = \forall B_j
  \forall A$.

\end{itemize}

$\blacksquare$

\noindent \textbf{Remark}.\label{remark:nonCommutatutivity} Operators
$\mathfrak{C}$ and $\mathfrak{D}$ commute, i.e. $\mathfrak{C} \circ
\mathfrak{D} = \mathfrak{D} \circ \mathfrak{C}$. In the equation of
item 4, the application order does not matter. Composition
$\mathfrak{D} \circ \mathfrak{C}$ is a direct translation of $\forall
A [\overline{A}]$ , which first considers all appearances of nodes in
$A$ and then splits these occurrences into separate digraphs. This is
the same as considering every pair of connected nodes in $A$ by one
edge and take their closure, i.e, $\mathfrak{C} \circ \mathfrak{D}$.

\begin{figure}[htbp]
  \centering
  \includegraphics[scale = 0.45]{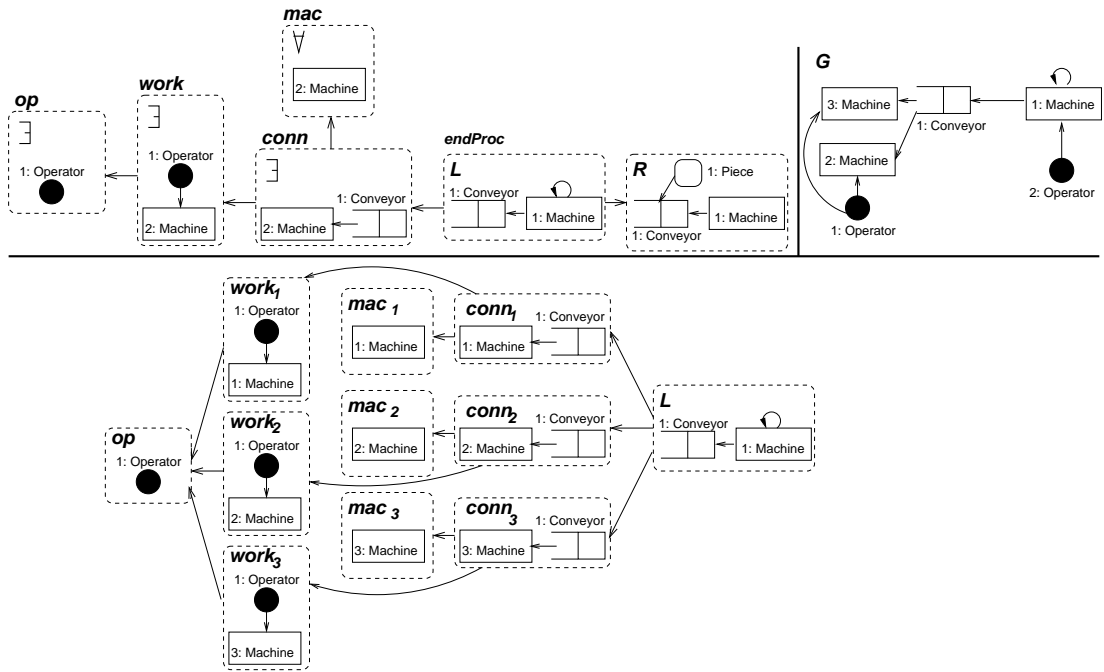}
  \caption{Example of Closure of AC with
    $depth>1$}\label{fig:example_closure_AC}
\end{figure}

\noindent {\bf Example.} Fig.~\ref{fig:example_closure_AC} shows
rule {\em endProc} and the diagram of its AC, which has formula:
$\exists op \: \forall mac$ $\exists work \: \exists conn [ (mac
\wedge conn) \Rightarrow (op \wedge work) ]$. The AC allows for the application of
the rule if all machines connected (as output) to the conveyor in $L$
are operated by the same operator.  This is so as the AC considers all
machines connected to the LHS conveyor by $\forall mac ...\exists
conn[mac \wedge conn]$.  For these machines, it should be the case
that a unique operator ($\exists op$ is placed at the beginning of the
formula) is connected to them ($\exists work$). 

The bottom of the figure shows the resulting diagram after applying
the previous theorem, using graph $G$ to the upper right of the
figure. At depth 2, graph $mac$ is replicated three times, as it is
universally quantified and there are three machines. Then, the rest of
the diagram is replicated, except the graphs quantified before $mac$
($L$ and $op$).  The resulting formula of the AC is $\exists op \:
\exists_{i=1}^3 mac_i \: \exists_{i=1}^3 work_i \: \exists_{i=1}^3
conn_i$ $[\bigwedge_{i=1}^3((mac_i \wedge conn_i) \Rightarrow (op \wedge
work_i) )]$, where we have omitted the $P_U$ predicate (asking that
actually three machines have to be found in $G$), and used the
abbreviation $\exists_{i=1}^3 A_i \equiv \exists A_1 \: \exists A_2 \:
\exists A_3$. Note that graph $G$ satisfies the AC (using the only
match of $L$ in $G$) as machine 1 is not operated by the same operator
as machines 2 and 3, however conveyor 1 is not connected to machine 1
as output (thus the left part of the implication is false).
$\blacksquare$

As an AC is a particular case of graph constraint, we can conclude
that it is not necessary to extend the notion of direct derivation in
order to consider ACs.

\begin{corollary}\label{cor:embedding}
  Any application condition $AC = \left( \mathfrak{d}, \mathfrak{f} =
    \mathfrak{f} \left( P, Q \right) \right)$ with $\mathfrak{f}$ a
  ground formula can be embedded into its corresponding direct
  derivation.
\end{corollary}

Now we are able to obtain ACs with existentials and total morphisms
only. The next section shows how to translate rules with such ACs into
sets of rule sequences.

One of the strengths of MGG compared to other graph transformation
approaches is the possibility to analyse grammars independently (to
some extent) of the actual host graph. However, the universal
quantifier appears to be an insurmountable obstacle: the host graph
seems indispensable to know how many instances there are. We will see
in section \ref{sec:proc_AC2Seq} that this is not the case.

\section{Transforming Application Conditions into Sequences}
\label{sec:AC_seq}
In this section we transform arbitrary ACs into sequences of plain
rules, such that if the original rule with ACs is applicable the
sequence is applicable and viceversa. This is very useful, as we may
use our analysis techniques for plain rules in order to analyse rules
with ACs.  Next, we present some properties of ACs which, once the AC
is translated into a sequence, can be analysed using the developed
theory for sequences.

\begin{definition}[Coherence, Compatibility,
  Consistency]\label{def:consistentAC}
  Let $AC = \left( \mathfrak{d}, \mathfrak{f} \right)$ be an AC on
  rule $p:L \rightarrow R$.  We say that AC is:
  \begin{itemize}
  \item \emph{coherent} if it is not a contradiction (i.e. false in
    all scenarios).
  \item \emph{compatible} if, together with the rule's actions,
    produces a simple digraph.
  \item \emph{consistent} if $\exists G$ host graph such that $G
    \models AC$ to which the production is applicable.
  \end{itemize}
\end{definition}

Coherence of ACs studies whether there are contradictions in it
preventing its application in any scenario. Typically, coherence is
not satisfied if the condition simultaneously asks for the existence
and non-existence of some element. Compatibility of ACs checks whether
there are conflicts between the AC and the rule's actions. Here we
have to check for example that if a graph of the $AC$ demands the
existence of some edge, then it can not be incident to a node that is
deleted by production $p$. Consistency is a kind of well-formedness of
the AC when a production is taken into account. Next, we show some
examples of non-compatible and non-coherent ACs.

\begin{figure}[htbp]
  \centering
  \includegraphics[scale = 0.5]{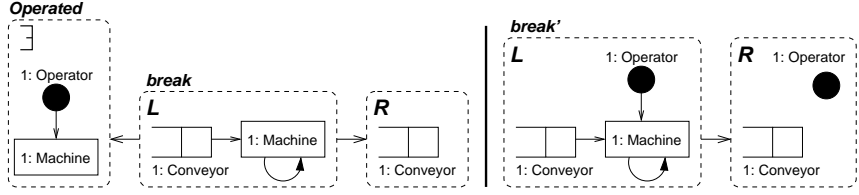}
  \caption{Non-Compatible Application
    Condition}\label{fig:example_non_compat}
\end{figure}

\noindent \textbf{Examples}. Non-compatibility can be avoided at times
just rephrasing the AC and the rule.  Consider the example to the left
of Fig.~\ref{fig:example_non_compat}. The rule models the breakdown of
a machine by deleting it. The AC states that the machine can be broken
if it is being operated. The AC has associated diagram $\mathfrak{d} =
\{Operated\}$ and formula $\mathfrak{f} = \exists Operated
[Operated]$. As the production deletes the machine and the AC asks for
the existence of an edge connecting the operator with the machine, it
is for sure that if the rule is applied we will obtain at least one
dangling edge.

\begin{wrapfigure}{r}{0.33\textwidth}
  \vspace{-0.5cm}
  \centering
  \includegraphics[width=0.31\textwidth]{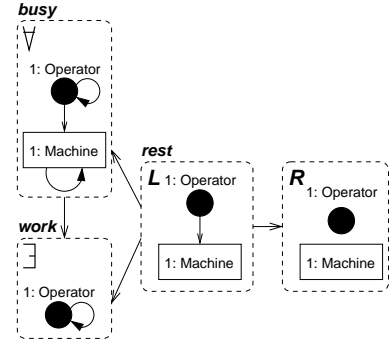}
  \caption{Non-Coherent AC.} \label{fig:example_non_coherent}
  \vspace{-0.8cm}
  \noindent
\end{wrapfigure}

The key point is that the AC asks for the existence of the edge but
the production demands its non-existence as it is included in the
nihilation matrix $N$.  In this case, the rule $break'$ depicted to
the right of the same figure is equivalent to $p$ but with no
potential compatibility issues.

Notice that coherence is fulfilled in the example to the left of
Fig.~\ref{fig:example_non_compat} (the AC alone does not encode any
contradiction) but not consistency as no host graph can satisfy it.

An example of non-coherent application condition can be found in
Fig.~\ref{fig:example_non_coherent}.  The AC has associated formula
$\mathfrak{f} = \forall busy \exists work [ busy \wedge$ $P ( work,
\overline{G}) ]$.  There is no problem with the edge deleted by the
rule, but with the self-loop of the operator.  Note that due to
$busy$, it must appear in any potential host graph but $work$ says
that it should not be present. $\blacksquare$

We will provide a means to study such properties by converting the AC
into a sequence of plain rules and studying the sequence, by applying
the analysis techniques already developed in MGG. We will prove that
an AC is coherent if its associated sequence is coherent and similarly
for compatibility. Also, we will see that an AC is consistent if its
associated sequence is applicable in some host graph. As this requires
sequences to be both coherent and compatible, and AC is consistent if
it is both coherent and
compatible~\cite{MGGCombinatorics}~\cite{MGGBook}.

\subsection{From ACs to Sequences: The Transformation Procedure}
\label{sec:proc_AC2Seq}

In order to transform a rule with ACs into sequences of plain rules,
operators $\mathfrak{C}$ and $\mathfrak{D}$ are expressed with the
bra-ket functional notation introduced in
definition~\ref{def:functional_rule}. Operators $\mathfrak{C}$ and
$\mathfrak{D}$ will be formally represented as $\widecheck{T}_{\!A}$
and $\widehat{T}_{\!A}$, respectively, and we analyse how they act on
productions and grammars. We shall follow a case by case study of the
demonstration of theorem \ref{th:embeddingGC} to structure this
section. The first case in the proof of theorem \ref{th:embeddingGC}
is the simplest one: a graph $A$ has to be found in $G$.

\begin{lemma}[Match]\label{lemma:firstCase}

  Let $p:L\rightarrow R$ be a rule with $AC = ( (A, d:L \rightarrow A
  ), \exists A [A])$, $p$ is applicable to graph $G$ iff sequence
  $p;id_{A}$ is applicable\footnote{Recall that sequence application
    order is from right to left.}  to $G$, where $id_{A}$ is a
  production with LHS and RHS equal to $A$.
\end{lemma}

\noindent {\em Proof.} The AC states that an additional graph $A$ has
to be found in the host graph, related to $L$ according to the
identifications in $d$. Therefore we can do the {\bf or} of $A$ and
$L$ (according to the identifications specified by $d$), and write the
resulting rule using the functional notation of
definition~\ref{def:functional_rule}, obtaining $\left\langle L \vee
  A, p \right\rangle$.  Thus applying the rule to its LHS, we obtain
$p(L \vee A) = R \vee A$.

Note however that such rule is the composition of the original rule
$p$, and rule $\abb{id_A}{A}{A}$. Thus, we can write $\left\langle L
  \vee A, p \right\rangle = \left\langle L, id_{\!A} \circ p
\right\rangle = p \circ id_{\!A}$, which proves also that
$id^*_{\!A}\!\left( L \right) = L \vee A$, the adjoint operator of
$id_{\!A}$. The symbol ``$\circ$'' denotes rule composition according
to the identification across rules specified by $d$
(see~\cite{JuanPP_1}). Thus, if the $AC$ asks for the existence of a
graph, it is possible to enlarge the rule $p \mapsto p \circ
id_{\!A}$. The marking operator $T_\mu$ permits using concatenation
instead of composition $\left\langle L \vee A, p \right\rangle = p ;
id_{\!A}$. $\blacksquare$

\begin{figure}[htbp]
  \centering
  \includegraphics[scale = 0.45]{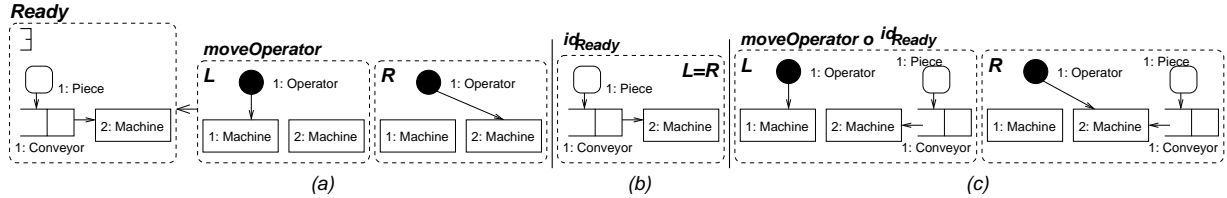}
  \caption{Transforming $\exists Ready[Ready]$ into a
    Sequence.}\label{fig:example_Exists}
\end{figure}

\noindent {\bf Example.} The AC of rule $moveOperator$ in
Fig.~\ref{fig:example_Exists} (a) has associated formula $\exists
Ready [Ready]$ (i.e. the operator may move to a machine with an
incoming piece).  Using previous construction, we obtain that the rule
is equivalent to sequence $moveOperator^{\flat}; id_{Ready}$, where
$moveOperator^{\flat}$ is the original rule without the AC. Rule
$id_{Ready}$ is shown in Fig.~\ref{fig:example_Exists} (b).
Alternatively, we could use composition to obtain
$moveOperator^{\flat} \circ id_{Ready}$ as shown in
Fig.~\ref{fig:example_Exists} (c). $\blacksquare$

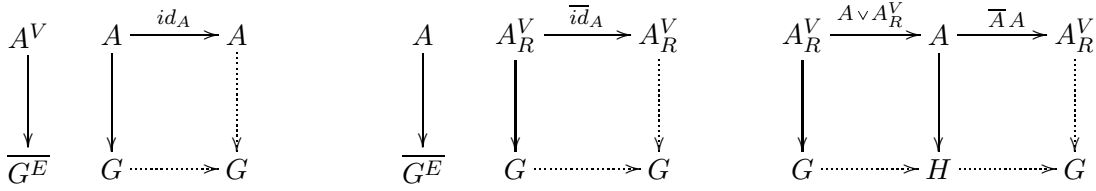
\begin{figure}[htb]
  \centering \makebox{ \xymatrix@C=0.47cm@R=0.47cm{
      A^V \ar[dd] & A \ar[dd] \ar[rr]^{id_A} && A \ar@{.>}[dd] &&& A \ar[dd] & A^V_R \ar[dd] \ar[rr]^{\overline{id}_A} && A^V_R \ar@{.>}[dd] && A^V_R \ar[dd] \ar[rr]^{A \vee A^V_R} && A \ar[dd] \ar[rr]^{\overline{A} \, A} && A^V_R \ar@{.>}[dd] \\ \\
      \overline{G^E} & G \ar@{.>}[rr] && G &&& \overline{G^E} & G
      \ar@{.>}[rr] && G & & G \ar@{.>}[rr] && H \ar@{.>}[rr] && G } }
  \caption{Identity $id_A$ (left), Conjugate $\overline{id}_A$ for
    Edges (center), $\overline{id}_A$ as Sequence for Edges (right).}
  \label{fig:idAndConj}
\end{figure}

The second case in the proof of theorem \ref{th:embeddingGC} states
that some edges of $A$ cannot be found in $G$ for some identification
of nodes in $G$, i.e. $\slash\!\!\forall A \left[ A \right] = \exists
A \left[ \overline{A} \right]$. This corresponds to operator
$\widehat{T}_{\!A}$ (decomposition), defined by
$\widehat{T}_{\!A}\left(p \right) = \left\{p_1, \ldots, p_n \right\}$.
For this purpose, we introduce a kind of conjugate (for edges) of
production $id_{A}$, written $\overline{id}_{A}$.  The left of
Fig.~\ref{fig:idAndConj} shows $id_A$, which preserves (uses but does
not delete) all elements of $A$. This is equivalent to demand their
existence.  In the center we have its conjugate, $\overline{id}_A$,
which asks for the existence of $A$ in the complement of $G$.

Rule $\overline{id}_A$ for edges can be defined on the basis of
already known concepts (i.e. having a ``normal'' nihilation matrix,
according to proposition~\ref{lemma:nihilMatrix}).  Since $N = r \vee
\overline{e} \, \overline{D}$, in order to obtain a rule applicable
iff $A^E$ is in $\overline {G^E}$, the only chance is to act on the
elements that some rule adds.  Let $p_e;p_r$ be a sequence such that
$p_r$ adds the edges whose presence is to be avoided and $p_e$ deletes
them.
The overall effect is the identity (no effect) but the sequence can be
applied iff the edges of $A$ are in $\overline{G^E}$ (see the right of
Fig.~\ref{fig:idAndConj}).  A similar construction does not work for
nodes because if a node is already present in the host graph a new one
can always be added (adding and deleting a node does not guarantee
that the node is not present in the host graph).  Thus, we restrict to
diagrams made of graphs without isolated nodes.  The way to proceed is
to care only about nodes that are present in the host graph as the
others together with their edges will be present in the completion of
the complement of $G$.  This is $A^V_R$, where $R$ stands for
\emph{restriction}.

Next lemma uses the previous conjugate rule to convert the ACs in the
second case of theorem~\ref{th:embeddingGC} into a set of rule
sequences.

\begin{lemma}[Decomposition]\label{lemma:secondCase}

  Let $p:L\rightarrow R$ be a rule with $AC = ( (A, d:L \rightarrow A
  ),\slash\!\!\forall A \left[ A \right])$, $p$ is applicable to graph
  $G$ iff some sequence in the set $\{s_i = p;\overline{id}_{A^i}\}$
  is applicable to graph $G$, with $\overline{id}_{A^i}$ the edge
  conjugate rule obtained from each graph $A^i$ in the decomposition
  of $A$.
\end{lemma}

\noindent {\em Proof.} Let $n$ be the number of edges of $A$, and
$A^i$ a graph consisting of one edge of $A$ (together with its source
and target nodes). Applying decomposition, the formula is transformed
into: $\mathfrak{f} = \exists A [\overline{A}] \longmapsto
\mathfrak{f}' = \exists A^1 \ldots \exists A^n \left[ \bigvee^n_{i=1}
  P \left( A^i, \overline{G} \right)\right]$. That is, the AC
indicates that more edges must not appear in order to apply the
production.  We build the set $\{p_i\}_{i \in \{1..n\}}$, where each
production $p_i$ is equal to $p$, but its nihilation matrix is
enlarged with $N_i = N \vee A^i$.  Thus, some production in this set
will be applicable iff some edge of $A$ is found in $\overline{G}$
(i.e. iff $\overline P(A, G)$ holds) and $p$ is applicable.  But note
that $p_i = p \circ \overline{id}_{\!A^i}$, where
$\overline{id}_{\!A^i}$ is depicted in the center of
Fig.~\ref{fig:idAndConj}.

If composition is chosen instead of concatenation, the grammar is
modified by removing rule $p$ and adding the set of productions
$\left\{p_1, \ldots, p_n \right\}$. If the production is part of a
sequence, say $q_2;p;q_1$ then we have to substitute it by some $p_i$,
i.e. $q_2;p;q_1 \mapsto q_2;p_i;q_1$. A similar reasoning applies if
we use concatenation instead of composition, where we have to replace
any sequence: $q_2;p;q_1 \mapsto q_2;p;\overline{id}_{A^i};q_1$, where
rules $p$ and $\overline{id}_{A^i}$ are related through
marking.$\blacksquare$

\begin{figure}[htbp]
  \centering
  \includegraphics[scale = 0.4]{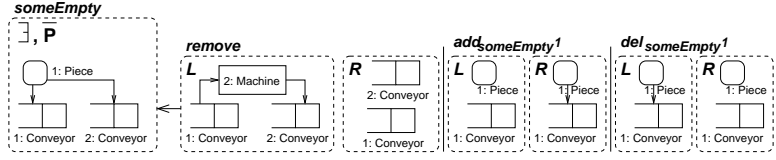}
  \caption{Transforming $\exists someEmpty[\overline {someEmpty}]$
    into a Sequence.}\label{fig:example_decomposition}
\end{figure}

\noindent {\bf Example.} The AC of rule $remove$ in
Fig.~\ref{fig:example_decomposition} has as associated formula
$\exists someEmpty [\overline {someEmpty}]$. The formula states that
the machine can be removed if there is one piece that is not connected
to the input or output conveyor (as we must not find a total morphism
from $someEmpty$ to $G$).  Applying the lemma~\ref{lemma:secondCase},
rule $remove$ is applicable if some of the sequences in the set $\{
remove^{\flat}; del_{someEmpty^i};$ $add_{someEmpty^i} \}_{i=\{1,
  2\}}$ is applicable, where productions $add_{someEmpty^2}$ and
$del_{someEmpty^2}$ are like the rules in the figure, but considering
conveyor 2. Thus $\overline {id}_{someEmpty^i}= del_{someEmpty^i}
\circ add_{someEmpty^i}$.$\blacksquare$

The third case demands that for any identification of nodes in the
host graph every edge must also be found: $\forall A\left[A\right] =
\nexists A [\overline{A}]$, associated to operator
$\widecheck{T}_{\!A}$ (closure).

\begin{lemma}[Closure]\label{lemma:thirdCase}

  Let $p:L\rightarrow R$ be a rule with $AC = ( (A, d:L \rightarrow A
  ),\forall A [A])$, $p$ is applicable to graph $G$ iff sequence
  $p;id_{\widecheck A}$ is applicable to graph $G$. $\widecheck A$ is
  the composition (through their common elements) of the graphs
  resulting from the closure of $A$ w.r.t. $G$.
\end{lemma}

\noindent {\em Proof.} Closure transforms $\mathfrak{f} = \forall A
[A] \longmapsto \exists A^1 \ldots \exists A^n$ $ \left[
  \bigwedge^n_{i=1} A^i \bigwedge_{i, j=1, j > i}^n P_U(A^i,
  A^j)\right]$, i.e. more edges must be present in order to apply the
production. Thus, we have to enlarge the rule's LHS: $L \longmapsto
\bigvee_{i=1}^n \left( L \vee A^i \right)$.  Using functional
notation, $\left\langle \bigvee_{i=1}^n \left(A^i \vee L \right), p
\right\rangle$ $=$ $\left\langle L,
  \widecheck{T_{\!A}}\!\left(p\right) \right\rangle$ $=$ $p \circ
id_{A^1} \circ \ldots \circ id_{A^n} = p \circ id_{\widecheck A}$, the
adjoint operator can be calculated as $\widecheck{T}^*_{A} \left( L
\right) = L \vee \left( \bigvee_{i=1}^n A^i \right)$.

As in previous cases, we may substitute composition with
concatenation: $\langle \bigvee_{i=1}^n (A^i \vee L),$ $ p \rangle = p
; id_{A^1} ; \ldots ; id_{A^n} = p ; id_{\widecheck A}$, where
$id_{\widecheck A} = id_{A^1} \circ \ldots \circ id_{A^n}$. Note
however that, if we use the expanded sequence (with $id_{A^i}$ instead
of $id_{\widecheck A}$) we have to make sure that each $id_{A^i}$ is
applied at each different instance. This can be done by defining a
marking operator similar to $T_{\mu}$.  $\blacksquare$

\noindent {\bf Remark.} Note that the result of closure depends on the
number and type of the nodes in the host graph $G$, which gives the
number of replicas of $A$ that have to be found.

\begin{figure}[htbp]
  \centering
  \includegraphics[scale = 0.4]{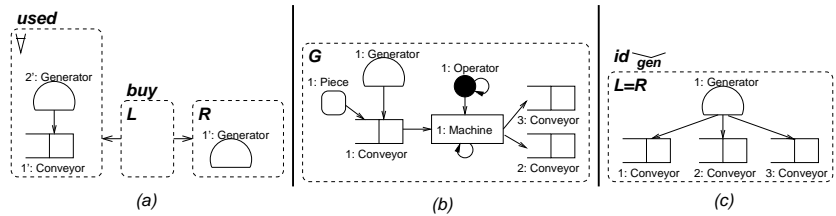}
  \caption{Transforming $\forall used[used]$ into a
    Sequence.}\label{fig:example_closure2}
\end{figure}

\noindent {\bf Example.} Fig.~\ref{fig:example_closure2} shows rule
{\em buy}, which creates a new generator machine. The rule has an AC
whose diagram is shown in the figure, with formula $\forall
used[used]$.  The AC permits applying the rule if all generators in the
host graph are connected to all conveyors.  Applying
lemma~\ref{lemma:thirdCase} to the previous rule and to graph $G$, we
obtain sequence $buy^{\flat}; id_{\widecheck{gen}}$.  As such sequence
is not applicable in $G$, the original rule is not applicable either.
$\blacksquare$.

The fourth case is in fact similar to a NAC, which is a mixture of (2)
and (3). 
This case says that there does not exist an identification of nodes of
$A$ for which all edges in $A$ can also be found, $\nexists A[A]$,
i.e. for every identification of nodes there is at least one edge in
$\overline{G^E}$.

\begin{lemma}[Negative AC]\label{lemma:nacs}

  Let $p:L\rightarrow R$ be a rule with $AC = ( (A, d:L \rightarrow A
  ),\not{\exists} A [A])$, $p$ is applicable iff some sequence
  $\widetilde{T_{A}}(p)=\left(\widehat{T}_{A} \circ \widecheck{T}_A
  \right)\! (p)$ is applicable.
\end{lemma}

\noindent {\em Proof.} Let $\widetilde{T_{A}}\! \left(p\right) =
\left(\widehat{T}_{A} \circ \widecheck{T}_{A} \right)\! (p) =
\left(\widecheck{T}_{A} \circ \widehat{T}_{A} \right)\! (p)$, then the
formula is transformed as follows: $\mathfrak{f} = \forall A
[\overline{A}] \longmapsto \exists A^{11} \ldots \exists$ $ A^{mn}
\left[ \bigwedge_{i=1}^m\bigvee_{\!j=1}^n A^{ij} \right]$.  If we
first apply closure to $A$ then we get a sequence of $m+1$
productions, $p \longmapsto p ; id_{A^1} ; \ldots ; id_{A^m}$,
assuming $m$ potential occurrences of $A$ in $G$.  Right afterwards,
decomposition splits every $A^i$ into its components (in this case
there are $n$ edges in $A$).  So every match of $A$ in $G$ is
transformed to look for at least one missing edge, $id_{A^1}
\longmapsto \overline{id}_{A^{11}} \vee \ldots \vee
\overline{id}_{A^{1n}}$.

Thus $\widetilde{T_A}(p)$ results in a set of rules
$\widetilde{T_{\!A}}\left(p \right) = \left\{p_1, \ldots, p_r
\right\}$ where $r = m^n$. Each $p_k$ is the composition of $m+1$
productions, defined as $p_k = p \circ \overline{id}_{A^{u_0v_0}}
\circ \ldots \circ \overline{id}_{A^{u_mv_m}}$. Operator $T_\mu$
permits concatenation instead of composition $\widetilde{T_A} (p) =
\left\{ p_k \; \vert \; p_k = p; \overline{id}_{A^{u_0v_0}} ; \ldots ;
  \overline{id}_{A^{u_mv_m}}\right\}_{k \in \{1, \ldots,
  m^n\}}$.$\blacksquare$

\begin{figure}[htbp]
  \centering
  \includegraphics[scale = 0.44]{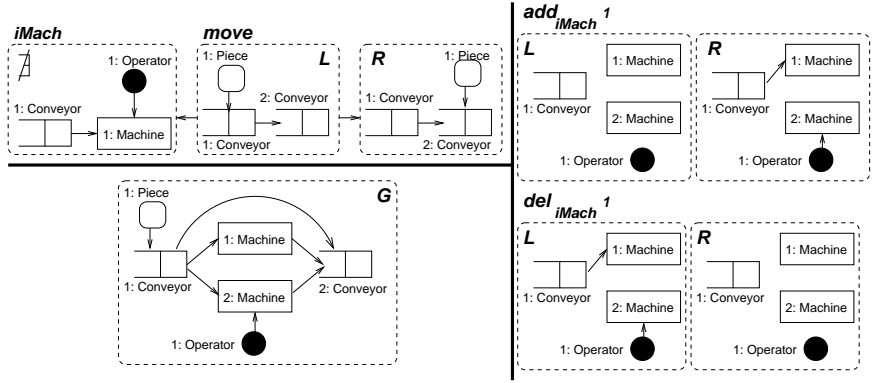}
  \caption{Transforming $\nexists iMatch[iMatch]$ into a
    Sequence.}\label{fig:example_NAC}
\end{figure}

\noindent {\bf Example.} Fig.~\ref{fig:example_NAC} shows rule
``move'' and a host graph $G$. A potential match identifies the
elements in $L$ with those in $G$ with the same number and type.  The
rule has an AC with associated formula $\nexists iMach[iMach]$.
Applying lemma~\ref{lemma:nacs}, we perform closure first, which
results in four potential instances of $iMach$:
$\{iMach^i\}_{i=1..4}$. Note however that only two of them preserve
the identification of elements given by the morphism $L \rightarrow
iMatch$ (as the conveyor in $L$ has to be matched to conveyor 1 in
$G$). The two instances contain the nodes $\{(1: Conveyor), (2:
Machine), (1: Operator)\}$ and $\{(1: Conveyor), (1: Machine), (1:
Operator)\}$ in $G$, the first contains in addition edges $(Operator,
Machine)$ and $(Conveyor, Machine)$, while the second contains the
$(Conveyor, Machine)$ edge only.

As each $iMach$ has two edges, decomposition leads to two rules for
each potential instance (each one detecting that one of the edges of
$iMach^i$ does not exist). Thus, we end up with 4 sequences of 3 rules
each (choosing concatenation of rules instead of composition). The
first two rules in each sequence detect that one edge is missing in
each potential instance of $iMatch$, while the last rule is
$move^{\flat}$. Note that choosing concatenation at this level makes
necessary a mechanism to control that each rule is applied at a
different potential instance of $iMach$. This is not necessary if we
compose these rules together.  The right of the figure shows one of
these compositions
($\overline{id}_{iMach^1}=\overline{id}_{iMach^{11}} \circ
\overline{id}_{iMach^{22}}$), which checks whether the first instance
of $iMach$ is missing the edge from the operator and the machine, and
the other one is missing the edge from the conveyor to the machine. As
before, we have split such rule in two:
$\overline{id}_{iMach^1}=del_{iMach^1}; add_{iMach^1}$. Thus,
altogether the applicability of the original rule move is equivalent
to the applicability of one of the sequences in $\{ move^{\flat};
del_{iMach^i}; add_{iMach^i} \}_{i=1..4}$, where each sequence can be
applied if each one of the two potential instances of $iMach$ is
missing at least one edge. Rules $move^{\flat}$ and $del_{iMach^i}$
are related through marking.  Note that none of these sequences is
applicable on $G$ (the first instance of $iMatch$ contains all edges),
thus the original rule is not applicable either. $\blacksquare$

Previous lemmas prove that ACs can be reduced to studying rule
sequences.

\begin{theorem}[Reduction of ACs]\label{th:reductionPre}

  Any AC can be reduced to the study of the corresponding set of
  sequences.
\end{theorem}

\noindent \emph{Proof} This result is the sequential version of
theorem \ref{th:embeddingGC}. The four cases of its proof correspond
to lemmas \ref{lemma:firstCase} through
\ref{lemma:nacs}.$\blacksquare$

\noindent {\bf Remark.} Quantifiers directly affect matching
morphisms. However, it is possible to some extent to apply all MGG
analysis techniques independently of the host graph, even in the
presence of universal quantifiers. The main idea is to
consider the initial digraph set (see
\cite{MGGBook}) of all possible starting graphs that enable the
sequence application. Some modifications of these graphs are needed to cope with
universals. The modified graphs in such set is then used to generate again
the sequences. Some examples of this procedure are given in section~\ref{sec:analysisAC}$\blacksquare$

\begin{figure}[htbp]
  \centering
  \includegraphics[scale = 0.44]{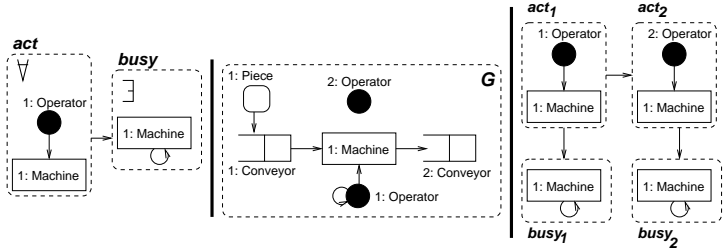}
  \caption{GC Example}\label{fig:final_example_GC}
\end{figure}

\noindent {\bf Example.} Fig.~\ref{fig:final_example_GC} shows a GC
with associated formula $\forall act \exists busy [act \Rightarrow
busy]$. The GC states that if an operator is connected to a machine,
such machine is busy.  Up to now we have focussed on analyzing ACs,
but the previous theorem also allows analyzing a GC as a set of
sequences. Note however that as the formula has an implication, it is
not possible to directly generate the set of sequences, as the GC is
also applicable if the left of the implication is false. Thus, the
easiest way is to apply the $\exists-P$ reduction of
theorem~\ref{th:embeddingGC}, which in this case reduces to applying
closure. The resulting diagram is shown to the right of the figure,
and the modified formula is then $\exists act_1 \exists act_2 \exists
busy_1 \exists busy_2 [ (act_1 \Rightarrow busy_1) \wedge (act_2
\Rightarrow busy_2)]$.

Once the formula has existentials only, we manipulate it to get rid of
implications. Thus, we have $\exists act_1 \exists act_2 \exists
busy_1 \exists busy_2 [(\overline{act_1} \vee busy_1) \wedge
(\overline{act_2} \vee busy_2)]= \exists act_1 \exists act_2 \exists
busy_1 \exists busy_2 [(\overline{act_1} \wedge \overline{act_2}) \vee
(\overline{act_1} \wedge busy_2) \vee (busy_1 \wedge \overline{act_2})
\vee (busy_1 \wedge busy_2)]$. This leads to a set of four sequences:
$\{(\overline{id}_{act_1}; \overline{id}_{act_2}),
(\overline{id}_{act_1}; id_{busy_2}), (id_{busy_1};
\overline{id}_{act_2}), (id_{busy_1}; id_{busy_2})\}$. Thus, graph $G$
satisfies the GC iff some sequence in the set is applicable to $G$.
However in this case none is applicable.

Testing GCs this way allows us checking whether applying a certain rule
$p$ preserves the GCs by testing the applicability of $p$ together
with the sequences derived from the GCs. This in fact gives
equivalent results to translating the GC into a post-condition for the
rule and then generating the sequences.  $\blacksquare$

\subsection{Analysing Graph Constraints and Application Conditions
  Through Sequences}
\label{sec:analysisAC}

As stated throughout the paper, one of the main points of the
techniques we have developed is to analyse rules with AC by
translating them into sequences of flat rules, and then analysing the
sequences of flat rules instead. In definition~\ref{def:consistentAC}
we presented some interesting properties to be analysed for ACs and
GCs (coherence, compatibility and consistency).  Next corollary, which
is a direct consequence of theorem~\ref{th:reductionPre}, deals with
coherence and compatibility of ACs and GCs.

\begin{corollary}\label{cor:equivPreAC_seqs}
  An AC is coherent iff if its associated sequence (set of sequences)
  is coherent; it is compatible iff its sequence (set of sequences) is
  compatible and it is consistent iff its sequence (set of sequences)
  is applicable.
\end{corollary}

In~\cite{JuanPP_4} (theorem 5.5.1) we characterized sequence
applicability as sequence coherence (see section~5
in~\cite{MGGCombinatorics} or section~4.3
in~\cite{JuanPP_4}) and compatibility (see section~4 and~7
in~\cite{MGGCombinatorics} or section 4.5
in~\cite{JuanPP_4}). Thus, we can state the following corollary.

\begin{corollary}\label{cor:precondConsCohComp}
  An AC is consistent iff it is coherent and compatible.
\end{corollary}

\noindent {\bf Examples.} Compatibility for ACs tells us whether there
is a conflict between an AC and the rule's action. As stated in
corollary~\ref{cor:equivPreAC_seqs}, this property is studied by
analysing the compatibility of the resulting sequence.  Rule {\em
  break} in Fig.~\ref{fig:example_non_compat} has an AC with formula
$\exists Operated[Operated]$.  This results in sequence:
$break^{\flat}; id_{Operated}$, where the machine in both rules is
identified (i.e. has to be the same).  Our analysis technique for
compatibility~\cite{JuanPP_1} outputs a matrix with a $1$ in the
position corresponding to edge $(1:Operator, 1: Machine)$, thus
signaling the dangling edge.

Coherence detects conflicts between the graphs of the AC (which
includes $L$ and $N$) and we can study it by analysing coherence of
the resulting sequence.  For the case of rule ``rest'' in Fig.~\ref{fig:example_non_coherent}, we would
obtain a number of sequences, each testing that ``busy'' is found, but
the self-loop of ``work'' is not. This is not possible, because this
self-loop is also part of ``busy''. Our technique for coherence
detects such conflict and the problematic element. $\blacksquare$

In addition, we can also use other techniques we have developed to
analyse ACs:

\begin{itemize}
\item {\bf Sequential Independence.} We can use our results for
  sequential independence of sequences to investigate if, once several
  rules with ACs are translated into sequences, we can for example
  delay all the rules checking the AC constraints to the end of the
  sequence. Note that usually, when transforming an AC into a
  sequence, the original flat rule should be applied last. Sequential
  independence allows us to choose some other order.  Moreover, for a
  given sequence of productions, ACs are to some extent delocalized in
  the sequence. In particular it could be possible to pass conditions
  from one production to others inside a sequence (paying due
  attention to compatibility and coherence). For example, a
  post-condition for $p_1$ in the sequence $p_2; p_1$ might be
  translated into a pre-condition for $p_2$, and viceversa.
\end{itemize}

\noindent {\bf Example.} The sequence resulting from the rule in
Fig.~\ref{fig:example_Exists} is $moveOperator^{\flat}; id_{Ready}$.
In this case, both rules are independent and can be applied in any
order. This is due to the fact that the rule effects do not affect the
AC. $\blacksquare$

\begin{itemize}
\item {\bf Minimal Initial Digraph and Negative Initial Digraphs}. The
  concepts of MID and NID allow us to obtain the (set of) minimal
  graph(s) able to satisfy a given GC (or AC), or to obtain the (set
  of) minimal graph(s) which cannot be found in $G$ for the GC (or AC)
  to be applicable. In case the AC results in a single sequence, we
  can obtain a minimal graph; if we obtain a set of sequences, we get
  a set of minimal graphs. In case universal quantifiers are present,
  we have to complete all existing partial matches so it might be
  useful to limit the number of nodes in the host graph under
  study.\footnote{This, in many cases, arises naturally. For example,
    in~\cite{MGGmodel} MGG is studied as a model of computation and a
    formal grammar, and also it is compared to Turing machines and
    Boolean Circuits. Recall that Boolean Circuits have fixed input
    variables, giving rise to MGGs with a fixed number of nodes. In
    fact, something similar happens when modeling Turing machines,
    giving rise to so-called (MGG) nodeless model of computation.}

  A direct application of the MID/NID technique allows us to solve the
  problem of finding a graph that satisfies a given AC. The technique
  can be extended to cope with more general GCs.
\end{itemize}

\begin{wrapfigure}{r}{0.38\textwidth}
  \vspace{-0.5cm}
  \centering
  \includegraphics[width=0.38\textwidth]{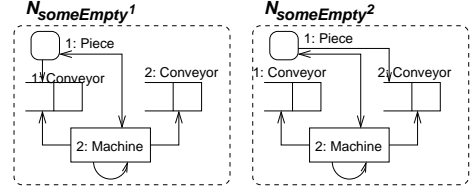}
  \caption{Negative Graphs Disabling the Sequences in
    Fig.~\ref{fig:example_decomposition}}\label{fig:example_NID}
  \vspace{-0.5cm}
  \noindent 
\end{wrapfigure}
  \noindent {\bf Example.} Rule $remove$ in
  Fig.~\ref{fig:example_decomposition} results in two sequences. In
  this case, the minimal initial digraph enabling the applicability
  for both is equal to the LHS of the rule.  The two negative initial
  digraphs are shown in Fig.~\ref{fig:example_NID} (and both assume a
  single piece in $G$). This means that the rule is not applicable if
  $G$ has any edge stemming from the machine, or two edges stemming
  from the piece to the two conveyors. $\blacksquare$

\begin{figure}[htbp]
  \centering
  \includegraphics[scale = 0.44]{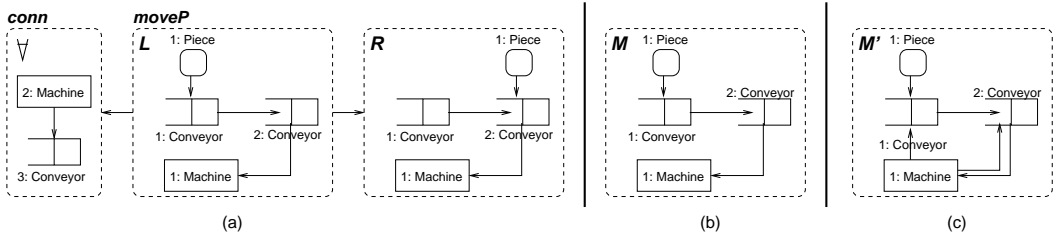}
  \caption{Completion of Minimal Digraph. (a) Example rule. (b)
    Minimal Digraph for Rule without AC. (c) Completed Minimal
    Digraph.}
  \label{fig:MID_Completion}
\end{figure}

  \noindent {\bf Example.} Fig.~\ref{fig:MID_Completion} shows the
  minimal initial digraph for executing rule $moveP$. As the rule has
  a universally quantified condition ($\forall conn[conn]$), we have
  to complete the two partial matches of the initial digraph so as to
  enable the execution of the rule.$\blacksquare$

\begin{itemize}
\item \textbf{G-congruence.} Graph congruence characterizes sequences
  with the same initial digraph. Therefore, it can be used to study
  when two GCs/ACs are equivalent for all morphisms or for some of
  them. See section~7 in~\cite{MGGCombinatorics} or section~6.1
  in~\cite{MGGBook}.
\end{itemize}

Moreover, we can use our techniques to analyse properties which up to
now have been analysed either without ACs or with NACs, but not with
arbitrary ACs:

\begin{itemize}
\item {\bf Critical Pairs.} A critical pair is a minimal graph in
  which two rules are applicable, and applying one disables the
  other~\cite{Heckel}. Critical pairs have been studied for rules
  without ACs~\cite{Heckel} or for rules with NACs~\cite{Lambers}. Our
  techniques however enable the study of critical pairs with any kind
  of AC. This can be done by converting the rules into sequences,
  calculating the graphs which enable the application of both
  sequences, and then
  checking whether the application of a sequence disables the other.

  In order to calculate the graphs enabling both sequences, we derive
  the minimal digraph set for each sequence as described in previous
  item. Then, we calculate the graphs enabling both sequences (which
  now do not have to be minimal, but we should have jointly surjective
  matches from the LHS of both rules) by identifying the nodes in each
  minimal graph of each set in every possible way. Due to universals,
  some of the obtained graphs may not enable the application of some
  sequence. The way to proceed is to complete the partial matches of
  the universally quantified graphs, so as to make the sequence
  applicable.

  Once we have the set of starting graphs, we take each one of them
  and apply one sequence. Then, the sequence for the second rule is
  recomputed -- as the graph has changed -- and applied to the
  graph. If it can be applied, there are no conflicts for the given
  initial graph, otherwise there is a conflict.  Besides the conflicts
  known for rules without ACs or with NACs (delete-use and
  produce-forbid~\cite{graGraBook}), our ACs may produce additional
  kinds of conflicts. For example, a rule can create elements which
  produce a partial match for a universally quantified constraint in
  another AC, thus making the latter sequence unapplicable. Further
  investigation on the issue of critical pairs is left for future
  work.
\end{itemize}

\begin{figure}[htbp]
  \centering
  \includegraphics[scale = 0.44]{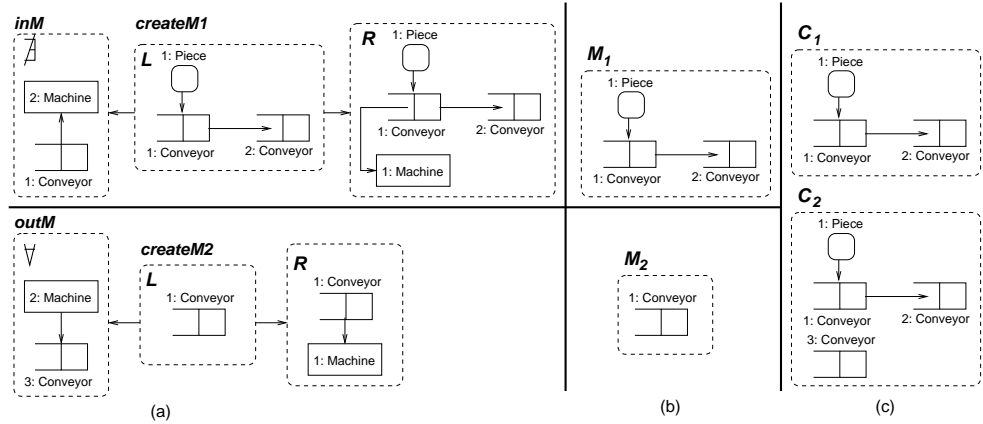}
  \caption{Calculating Critical Pairs. (a) Example Rules. (b) Minimal
    Digraphs. (c) Starting Graphs for Analysing
    Conflicts.}\label{fig:Critical_Pairs}
\end{figure}

\noindent {\bf Example.} Fig.~\ref{fig:Critical_Pairs}(a) shows two
rules, $createM1$ and $createM2$, with ACs $\nexists inM[inM]$ and
$\forall outM[outM]$ respectively. The center of the same figure
depicts the minimal digraphs $M_1$ and $M_2$, enabling the execution
of the sequences derived from $createM1$ and $createM2$
respectively. In this case, both are equal to the LHS of each rule.
The right of the figure shows the two resulting graphs once we
identify the nodes in $M_1$ and $M_2$ in each possible way. These are
the starting graphs that are used to analyse the conflicts.  The rules
present several conflicts. First, rule $createM1$ disables the
execution of $createM2$, as the former creates a new machine, which is
not connected to all conveyors, thus disabling the $\forall
outM[outM]$ condition of $createM2$.  The conflict is detected by
executing the sequence associated to $createM1$ (starting from either
$C_1$ or $C_2$), and then recomputing the sequence for $createM2$,
taking the modified graph as the starting one. Similarly, executing
rule $createM2$ may disable $createM1$ if the new machine is created
in the conveyor with the piece (this is a produce-forbid
conflict~\cite{Lambers}). $\blacksquare$

\begin{itemize}
\item {\bf Rule Independence}. Similarly, results for rule
  independence have been stated either for plain rules, or rules with
  NACs. In our case, we convert the rules into sets of sequences and
  then check each combination of sequences of the two rules.
\end{itemize}

\section{Discussion and Comparison with Related Work}
\label{sec:related}

In the categorical approach to graph transformation,
ACs~\cite{AC:Ehrig} are usually defined by Boolean formulae of
positive or negative atomic ACs on the rule's LHS. The atomic ACs are
of the form $P(x, \bigvee_{i \in I} x_i)$ or $N(x, \bigwedge_{i \in I}
x_i)$, with $\abb{x}{L}{X}$ and $\abb{x_i}{X}{C_i}$ total functions.
The diagrams in this kind of ACs are limited to depth 2 and there is
no explicit control on the quantifications. In our approach, the ACs
are not limited to be constraints on the LHS, thus we can use
``global'' information, as seen in the examples of
Figs.~\ref{fig:example_AC} and~\ref{fig:example_closure_AC}. This is
useful for instance to state that a certain unique pattern in the host
graph is related to all instantiations of a certain graph in the AC.
Moreover, in our ACs, the diagrams may have any shape (and in
particular are not limited to depth 2).  Whether elements should be
mapped differently or not is tackled by restricting the morphisms from
the ACs to the host graph to be injective in~\cite{HP06}. On the
contrary, we use partial functions and predicate $P_U$.  Our use of
the closure operator takes information from the host graph and stores
it in the rule. This enables the generation of plain rules, whose
analysis is equivalent to the analysis of the original rule with ACs.

In~\cite{Habel}, the previous concept of GCs and ACs were extended
with nesting. However, their diagrams are still restricted to be
linear (which produces tree-like ACs), and quantification is performed
on the morphisms of the AC (i.e. not given in a separate formula).
Again, this fact difficults expressing ACs like those in
Figs.~\ref{fig:example_AC} and~\ref{fig:example_closure_AC}, where a
unique element has to be related to all instances of a given graph,
which in its turn have to be related to the rule's LHS. In~\cite{HP09},
the same authors present techniques for transforming graph constraints into right
application conditions and those to pre-conditions, show the equivalence of
considering non-injective and injective matchings, and the equivalence of
GCs and first order graph-formulae. The work is targeted to the verification
of graph transformation systems relative to graph constraints (i.e., to check whether
the rules preserve the constraints or not, or to derive pre-conditions ensuring that
the constraints are preserved). In our case,
we are interested in analysing the rules themselves (see Section~\ref{sec:analysisAC}),
e.g. checking independence, or calculating the minimal graph able to fire a sequence
using the techniques already developed for plain rules. We have left out related topics,
such as the transformation from pre- to post-conditions, which are developed in the
doctoral thesis available at~\cite{MGGBook}.
Note however, that there are
some similarities between our work and that of~\cite{HP09}. For example, in their theorem
8, given a rule, they provide a construction to obtain a GC that if satisfied, permits
applying the rule at a certain match. Hence, the derived GC makes explicit the glueing
condition and serves a similar purpose as our nihilation matrix. Notice however that the
nihilation matrix contains negative information and has to be checked on the negation
of the graph.

The work of~\cite{Rensink} is an attempt to relate logic and algebraic
rewriting, where ACs are generalized to arbitrary levels of nesting
(in diagrams similar to ours, but restricted to be trees).
Translations of these ACs into first order logic and back are given,
as well as a procedure to flatten the ACs into a normal graph, using
edge inscriptions. We use arbitrary diagrams, complemented with a MSOL
formulae, which includes quantifications of the different graphs of
the diagram. Our goal was to flatten such ACs into sequences of plain
rules.

Related to the previous work, in~\cite{Canonical}, a logic based on
first-order predicate is proposed to restrict the shape of graphs. A
decidable fragment of it is given called local shape logic, on the
basis of a multiplicity algebra. A visual representation is devised
for monomorphic shapes. This approach is somehow different from ours,
as we break the constraint into a diagram of graphs, and then give a
separate formula with the quantification.

Thus, altogether, the advantages of our approach are the following:
(i) we have a universal quantifier, which means that some conditions
are more direct to express, for example taking the diagram of
Fig.~\ref{fig:example_fulfill1}, we can state $\forall bOp[bOp]$,
which demands a self-loop in all operators. In the algebraic approach
there is no universal quantifier, but it could be emulated by a
diagram made of two graphs stating that if an operator exists then it
must have a self-loop.  However, this becomes more complicated as the
graphs become more complex. For example, let $A$ be a graph with two
connected conveyors (in each direction). Then $\forall A[Q(A, G)]$
asks that each two conveyors have at least a connection. In the
algebraic approach, one has to take the nodes of $A$ and check their
existence, and then take each edge of $A$ and demand that one of them
should exist. Note that this universal quantifier is also different
from amalgamation approaches~\cite{Taentzer}, which, roughly, are used to build a match
using all occurrences of a subgraph. In our case, we in addition demand each partial
occurrence to be included in a total one. (ii) We have an explicit control of the formula and the
diagram, which means that we can use diagrams with arbitrary shape,
and we can put existentials before universals, as in the example of
Fig.~\ref{fig:example_AC}. Again, this facilitates expressing such
constraints with respect to approaches like~\cite{HP09}. (iii) Sequences of plain rules can be
automatically derived from rules with ACs, thus making uniform the
analysis of rules with ACs.

On the contrary, one may argue that our universal is ``too strong'' as
it demands that all possible occurrences of a given graph are actually
found. This in general presents no problems, as a common technique is
for example to look for all nodes of a given graph constraint with a
universal, and then look for the edges with existentials.

With respect to other similar approaches to MGGs, in~\cite{Valiente}
the DPO approach was implemented using Mathematica. In that work,
(simple) digraphs were represented by Boolean adjacency matrices. This
is the only similarity with our work, as our goal is to develop a
theory for (simple) graph rewriting based on Boolean matrix algebra.
Other somehow related work is the relational approaches
of~\cite{Kahl,Mizoguchi}, but they rely on category theory for
expressing the rewriting. Similar to our dynamic formulation of
production and to our deletion and addition matrices, the approach of
Fujaba~\cite{Fujaba} considers the LHS of a production and labels with
``new'' and ``del'' the elements to be created and deleted.  Finally,
it is worth mentioning the set-theoretic approaches to graph
transformation~\cite{Engelfriet,Raoult92}. Even though some of these
approaches have developed powerful analysis techniques and efficient
tool implementations, the rewriting is usually limited (e.g. a node or
edge can be replaced by a subgraph).

\section{Conclusions and Future Work}
\label{sec:conclusions}

We have presented a novel concept of GCs and ACs based on a diagram of
graphs and morphisms and a MSOL formulae.  The concept has been
incorporated into our MGG framework, which in addition has been
improved by incorporating the notion of nihilation matrix. This matrix
contains edges that if present forbid rule application.  One
interesting point of the introduced notion of AC is that it is
possible to transform them into a sequence of plain rules, with the
same applicability constraints as the original rule with ACs. Thus, in
MGG we can use the same analysis techniques for plain rules and rules
with ACs.

We have left out some related topics, such as
post-conditions and transformation from pre- to post-conditions and
viceversa, the handling of nodes with variable type (i.e. nodes that
in the AC can get matched to nodes with other type in the host graph)
and its relation to meta-modelling~\cite{MGGBook}. This notion of ACs enables
performing multi-graph rewriting with simple graph rewriting by
representing edges as special nodes, plus a set of ACs.  Thus, MGG can
handle multigraphs with no further modification of the theory.

As future work, we are developing a tool implementation of the MGG
framework, enabling interoperability with existing graph grammars
tools such as AToM$^3$~\cite{AToM3} or AGG~\cite{AGG}.  We also plan
to include more complex means for typing (like a type graph) and
attributes in our framework.  Defining more general ACs, whose graphs
are not restricted to be connected, is also under consideration.
Following the ideas in~\cite{RuleQuantif} it could also be interesting
to permit quantification on rules themselves (and not only the ACs).
We also plan to deepen in the analysis of critical pairs, especially
analysing the new kind of conflicts arising due to our ACs, as well as by using the
negative initial digraphs for the analysis.

Finally, the presented concepts of GC and AC could be integrated with
other approaches to graph transformation, like the algebraic one.
There are some issues though, that cannot be directly translated into DPO/SPO: we
use the negation of a graph, and work with simple digraphs, which have
the built-in restriction that between two nodes at most one edge in
each direction is allowed.


\end{document}